%% file: 000-Main.tex
\documentclass[letterpaper]{article} 
\usepackage{aaai25}  
\usepackage{times}  
\usepackage{helvet}  
\usepackage{courier}  
\usepackage[hyphens]{url}  
\usepackage{graphicx} 
\usepackage{natbib}  
\usepackage{caption} 
\usepackage{algorithm}
\usepackage{algorithmic}

\usepackage[title]{appendix}

\usepackage{nicematrix}

\usepackage{newfloat}
\usepackage{listings}
\DeclareCaptionStyle{ruled}{labelfont=normalfont,labelsep=colon,strut=off} 
\floatstyle{ruled}
\newfloat{listing}{tb}{lst}{}
\floatname{listing}{Listing}
\renewenvironment{quote}
  {\list{}{\rightmargin=0.1em\leftmargin=0.1em}%
   \item\relax}
  {\endlist}

\title{Can LLM Code Explanations Adapt to Diverse Problem-Solvers' Needs?}
\author {
    Andrew Anderson\textsuperscript{\rm 1},
    David Piorkowski\textsuperscript{\rm 2},
    Justin D. Weisz\textsuperscript{\rm 1},
    Margaret Burnett\textsuperscript{\rm 3},
    Kush R. Varshney\textsuperscript{\rm 1}
}
\affiliations {
    \textsuperscript{\rm 1}IBM Research\\
    \textsuperscript{\rm 2}Alinia AI\\
    \textsuperscript{\rm 3}Oregon State University\\
    Andrew.Anderson2@ibm.com, 
    david@alinia.ai, 
    jweisz@gmail.com,
    burnett@eecs.oregonstate.edu,
    krvarshn@us.ibm.com
}

\usepackage{soul}
\usepackage{enumitem}
\usepackage{rotating}
\usepackage{multirow}
\usepackage{colortbl}
\usepackage{xcolor}
\usepackage{tcolorbox}
\usepackage{diagbox}
\usepackage{siunitx}
\usepackage{subcaption}
\usepackage{textcomp}
\usepackage{mathtools}
\usepackage{amsmath}
\usepackage{adjustbox}
\usepackage{listings}
\usepackage{xstring}
\usepackage{pdflscape}
\usepackage{rotating}
\usepackage{slashbox}
\usepackage{longtable}
\usepackage{booktabs}
\usepackage{transparent}
\usepackage{afterpage}
\usepackage{tabularx}
\usepackage{amsthm}
\theoremstyle{definition}
\newtheorem*{definition*}{Definition}

\input{z-commands-imports-options/00_aa_preamble_statements}

\begin{document}
\maketitle
\begin{abstract}
Large language model (LLM) code explanations can support people in solving code-related problems, yet prior work has shown that people have diverse problem-solving styles.
If explanations fail to meet people's problem-solving needs, they may be less productive in their occupations and miss opportunities to learn and grow. 
Although some research has examined how LLMs can adapt their outputs to a user's age or expertise, no prior work has examined how LLMs can adapt their code explanations to people's problem-solving styles.
To address this gap, we developed prompts from an established inclusive design method that considers 5 types of problem-solving styles, and we generated 1,072 code explanations from six open-weight LLMs.
Using natural language processing techniques, we uncovered a taxonomy of 13 linguistic adaptations, with each adaptation supported by evidence from the literature, the prompts, or the LLMs' outputs.
They also show which LLMs adapted their code explanations more frequently than others.
This paper is the first to investigate problem-solving style adaptations in LLM code explanation, contributing two problem-solving adaptation approaches: declarative statements for each adaptation and 10 problem-solving style prompts.
\end{abstract}

\section{Introduction}


\boldify{LLMs help people with code, but it's important that they understand it}

Large language models (LLMs) have shown capabilities in generating code~\cite{becker2023generative,liang2024large}, documenting it~\cite{feng2020codebert,luo2024repoagent}, and testing it~\cite{schafer2023empirical}, thereby mitigating technical debt~\cite{ghammam2025build}. 
However, people still need to \textit{understand} the code, in order to do software engineering work like debugging or maintenance~\cite{weisz2025examining}.

\boldify{LLMs could help with code explanations, but explanations aren't one-size-fits-all.}

LLMs can help with code understanding by explaining it, but prior research has found that explanations may not work person-to-person;
people have diverse needs for information presentations~\cite{Anik2024ExplPresentation}, granularity~\cite{Chatti2022ExplDepth}, or even types for explanation~\cite{dodge-fairness-2019}, depending on the situation~\cite{Tsakalakis2025ExplDesign}.
In essence, few explanations are ``one-size-fits-all''~\cite{anderson2019explaining,anderson2020mental,arya2020aix,dodge2021no}.

\boldify{However, all is not lost; LLMs can adapt, but the question becomes ``adapt to what?''}

\boldify{The question then becomes how? Problem-solving styles may point out how.}

The advances in LLMs' capabilities to dynamically adapt their outputs open opportunities to address this issue.
Prior work has shown how LLMs adapt their outputs to align with various diversity dimensions, like age \cite{Murgia2023AgeAdapt,rooein2023know}, education level \cite{Haver2023EduAdapt,rooein2023know}, and domain expertise \cite{Asthana2024DomainAdapt,Guo2024JargonAdapt}. 

Still, when
explanations aim to support problem solving~\cite{Chen2025ExplProbSolve}, as with software development, a more direct path to supporting diverse problem-solvers is to adapt LLMs' code explanations  directly to people's diverse problem-solving approaches~\cite{anderson-diversity-2024,anderson2026over-the-hood,hamid2024inclusive,nam2024using}. 
However, little is known about whether (and how) LLMs adapt their code explanations to people's diverse problem-solving styles. 
This paper fills that gap, answering a central research question:

\begin{itemize}[ label = {}, leftmargin = 1.2cm]
    \item[\textbf{RQ}:] How do LLMs adapt their code explanations to diverse \probType{s}?
\end{itemize}


\boldify{In this paper, we investigate how LLMs might adapt their code explanations to people's diverse problem-solving styles}

To answer this question, we focused on five \probType{s} from the Gender Inclusiveness Magnifier (GenderMag) method \cite{burnett2016gendermag}, namely: (1) learning style, (2) self-efficacy, (3) risk-attitude, (4) motivations, and (5) information processing style.
Each type has a range of continuous values, but prior work emphasizes the two endpoints for these types, giving us 10 \probVal{s} to consider.

\boldify{To address this question, we experimented with six open-source LLMs, generating and analyzing 1,072 explanations of three programs.}

With these 10 \probVal{s}, we developed an explanation generation and analysis pipeline.
For each \probVal{}, we developed a prompt from GenderMag's validated problem-solving style survey~\cite{hamid-2023}, and included an 11th experimental control.
We queried six open-source LLMs to explain Common Business-Oriented Language (COBOL) programs, an ongoing area for explanation research~\cite{lei2025enhancing}.
Using zero-shot prompting, we analyzed 1,072 code explanations for linguistic adaptations with standard natural language processing (NLP) and statistical methods.

Our work makes the following contributions:
\begin{itemize}
    \item Empirically validated problem-solving style prompts to elicit adaptations.
    \item A taxonomy of 13 problem-solving style adaptations, supported by the literature.
    \item An empirical ranking of LLMs by their frequency of code explanation adaptations to these  problem-solving styles.
\end{itemize}

\begin{table}[b]
    \small
    \centering
    \begin{tabularx}{\columnwidth}{p{0.27\columnwidth}|p{0.22\columnwidth}cp{0.22\columnwidth}}
        \toprule
        \textbf{Problem-Solving Style Type} &  \textbf{Extreme Value 1} & \multirow{2}{*}{$\iff$} & \textbf{Extreme Value 2} \\
        \midrule
        \multirow{2}{*}{\textbf{Learning Style}} & \cellcolor{abiorange}{Process-Oriented}&
        \multirow{2}{*}{$\iff$}&
        \cellcolor{timblue}{Tinkering-Oriented}
            \\
        \midrule
        
        \textbf{Self-Efficacy} & \cellcolor{abiorange}{{Lower}}&
        {$\iff$}&
        \cellcolor{timblue}{Higher}
            \\
        \midrule
        
        \textbf{Risk Attitudes} &
        \cellcolor{abiorange}{{Risk-Averse}}&
        {$\iff$}&
        \cellcolor{timblue}{Risk-Tolerant}
            \\
        \midrule

        \textbf{Motivations} &
            \cellcolor{abiorange}{{Task-Oriented}}&
         {$\iff$}&
        \cellcolor{timblue}{Tech-Oriented}
            \\
        \midrule
        
        \textbf{Information Processing Style} &
        \cellcolor{abiorange}\multirow{2}{*}{{Comprehensive}}&
         {$\iff$}&
        \cellcolor{timblue}\multirow{2}{*}{{Selective}}
            \\

        \bottomrule
    \end{tabularx}
    \caption{GenderMag's five \probType{s} (row), and the extreme \probVal{s} at the endpoints of each type (columns).
    We bind two colors to distinguish each \probVal{} endpoint.
    }
    \label{tab:01_gm_persona_table}
\end{table}

\section{Background \& Related Work}
\label{subsec:background}

\paragraph{The Gender Inclusiveness Magnifier (GenderMag)}

\boldify{A key underpinning of our work is GenderMag, which might not be familiar with a lot of people}

GenderMag's five \probType{s} underpin our prompting approach, so we provide an introductory background for each type%
\footnote{\citet{stumpf2020gender} thoroughly review the five types.}
(Table~\ref{tab:01_gm_persona_table}, rows).
Each type has a continuous spectrum of \textit{\probVal{s}}, with two extreme value endpoints (columns).
As \citet{fallatah2025intersectionalHCI} explain, designing for both endpoints also supports people whose \probVal{s} lie between the endpoints.

\boldify{The first \probType{} is learning style, which deals with how people like to learn about new technology}

\textbf{Learning Style} considers how people prefer to learn about new technology.
\personaVal{abiorange}{Process-oriented} learners prefer taking a structured approach, like following tutorials, manuals, and how-to videos.
\personaVal{timblue}{Tinkering-oriented} learners tend towards constructing their own understanding, exploring menus and software functions on their own~\cite{Hadad2025LearnStyle}.

\textbf{Self-Efficacy}, a type of self-confidence, considers a person's belief that they can succeed at tasks~\cite{bandura1986explanatory}.
When something goes wrong, those with \personaVal{abiorange}{lower} self-efficacy are more likely to blame themselves, reducing how likely they are to persevere.
Those with \personaVal{timblue}{higher} self-efficacy are more likely to blame the software, persevering when faced with adversity~\cite{sakdavong2025gender}.

\textbf{Risk Attitudes} considers how people engage with risk, including the risk of wasting time.
While making cost/benefit analyses, \personaVal{abiorange}{risk-averse} people may emphasize the potential costs, especially with unclear benefits.
By contrast, \personaVal{timblue}{risk-tolerant} people may emphasize the potential benefits~\cite{Kramer2025RiskAversionCostBenefit,Li2026RiskAversionCostBenefit}

\textbf{Motivations} considers the reasons why people interact with technology.
Those with \personaVal{abiorange}{task-oriented} motivations use technologies to accomplish their tasks.
They prefer methods they know and are comfortable with.
Those with \personaVal{timblue}{tech-oriented} motivations view technology as a source of fun.
They learn about all available functionalities, even those not necessary to their task~\cite{burnett2011gender}.

\textbf{Information Processing Style} considers how people gather information while solving problems.
People with a \personaVal{abiorange}{comprehensive} style take a ``breadth first'' approach, aiming towards a complete understanding of the problem before solving it.
\personaVal{timblue}{Selective} processors take a ``depth first'' approach, delving into the first promising option and backtracking to find another if necessary~\cite{ouyang2025female}.

\boldify{LLMs can explain code, but nobody agrees on their content...}

\paragraph{Related Works: LLM Code Explanations}
Prior work on LLM-generated code explanations has focused on explaining fundamental concepts like time complexity, the presence of errors, or the purpose of code~\cite{macneil2022generating}.
Other works have focused on how to incorporate examples and real-world analogies into code explanations~\cite{brachman2025towards}.



\boldify{...or their granularity}

Code explanation research also covers a broad range of code explanation granularities.
Some works have focused on how to generate concise summaries of individual programs and repositories~\cite{Liang2019FixExpl,Phillips2022CodeExplanation,Sun2025CodeExplAdapt,Sundaram2025CodeExpl}.
Other works have focused on more granular explanations, including line-by-line~\cite{Chapagain2024VerboseExpl}, function-level~\cite{leinonen2023comparing}, or user-specified selections~\cite{nam2024using}.

Our work focuses on COBOL code explanations, which others have considered for both LLM explanation content and granularity.
In particular, \citet{lei2025enhancing} evaluated LLM capabilities for function-level, file-level, and project-level explanations.
Additionally, they evaluated these explanations by their clarity, conciseness, and correctness.

These works provide insight into the myriad ways that code can be explained.
The focus of this paper is less on specific explanation dimensions (e.g., content, granularity) and instead considers whether (and how) LLMs adapt code explanations for GenderMag's \probType{s}.

\boldify{Although explanation engagement is problem-solving (e.g., bridging knowledge gaps), no prior work has examined whether LLMs can adapt to diverse problem-solving approaches.}



\paragraph{Related Works: LLM Adaptations}
\boldify{LLM adaptation research focuses on two broad areas, adopting personas and adapting to specific dimensions}

Some LLM adaptation works have examined how LLMs adopt a wide variety of personas, including professions~\cite{Ribeiro2026LLMJudgePersona} and demographic dimensions like age, sex, nationality, and religion~\cite{Tan2026CanLLMPersona}.
Conceptually close to our work, \citet{Geuenich2026PersonaLLM} asked GPT-4o-mini to \textit{become} the three GenderMag personas.
They measured how capable the LLM was in finding usability bugs that would disproportionately impact each persona.
This paper does not use GenderMag's personas, instead investigating how LLMs adapt code explanations to the five \probType{s}.

Other LLM adaptation research is on adapting to specific user dimensions.
One popular area focuses on demographic dimensions like age~\cite{Murgia2023AgeAdapt,rooein2023know}, education levels~\cite{Haver2023EduAdapt,rooein2023know}, and domain expertise~\cite{Asthana2024DomainAdapt,Guo2024JargonAdapt}.
Some of these works in code explanations have found that when LLMs are explaining to beginning learners, they generate significantly larger explanations~\cite{oli2023behavior}.

Other research has focused on adaptations for user dimensions similar to the five \probType{s} in this paper.
In particular, \citet{agarwal2025putting} investigated how GPT3.5 adapted its responses, using Kirton's Adaptation-Innovation Theory~\cite{kirton1984adaptors} and found successful adaptations for both innovators and adapters.
\citet{Harry2026PsychStateAdapt} examined how multiple models adapted to psychological state (e.g., anxious, distressed).
Their results uncovered that although the models recognized persona framings, the models failed to adapt to psychological states.
These varying success levels suggest that adaptations for GenderMag's five \probType{s} were plausible but not guaranteed.

\section{Methodology}
\label{sec:methods}

\paragraph{Problem-Solving Style Value Prompts}

\boldify{We needed to apply GenderMag to LLMs to adapt, and here's broad-strokes how.}

We developed 10 prompts from GenderMag's validated \probVal{} survey~\cite{hamid-2023} to steer the LLMs, each with \textit{introductory}, \textit{agreement}, and \textit{alignment} blocks.

The \textit{introductory block} introduced the task and minimally expected behavior.
This block was also the experimental control to produce unadapted explanation baselines.
Inspired by Meta's LLaMa~\cite{Meta2023LLaMaPrompt}, this block instructed the LLMs that they were a programmer's assistant and to not be unethical or harmful:
\begin{quote}
\footnotesize
    \textit{You are a programmer's assistant. 
    You can answer conceptual programming questions and explain what code samples do.
    Your responses are helpful and harmless and should follow ethical guidelines and promote positive behavior. 
    Your responses should not include unethical, racist, sexist, toxic, dangerous, or illegal content. 
    Ensure your responses are socially unbiased.}
\end{quote}

The \textit{agreement block} provided both the \probType{} and the specific value from the survey.
Each type was introduced in the following template:
\begin{quote}
    \footnotesize
    \textit{Research has shown that AI technology users have diverse $<<$\probType{}$>>$}.
\end{quote}
For each value, the prompt contained both the value name and the agreement statements.
To illustrate this, the following was used for \personaVal{abiorange}{process-oriented} users:
\begin{quote}
    \footnotesize
    \textit{Your user is a process-oriented learner, meaning they are more likely to disagree with the following statements:}.
    \begin{enumerate}[label = {\textit{\arabic*.}},itemsep=2pt, parsep = 0pt, topsep=0pt]
    \item \textit{I enjoy finding the lesser-known features and capabilities of the devices and software I use.}
    \item \textit{I don't follow instruction manuals. I only look to instruction manuals as a last resort.}
    \item \textit{I'm never satisfied with the default settings for my devices; I customize them in some way.}
    \item \textit{My first step in learning new technology is experimenting and tinkering with it.}
    \item \textit{I explore areas of a new application or service before it is time for me to use it.}
    \item \textit{I don't need guidance, as in booklets, video how-tos suggestions, etc. to learn new software.}
\end{enumerate}
\end{quote}
For the \personaVal{timblue}{tinkering-oriented} learning style prompt, the only changes to the agreement block were:
\begin{quote}
    \footnotesize
    \textit{Your user is a \textbf{tinkering-oriented} learner, meaning they are more likely to \textbf{agree} with the following statements:}
\end{quote}

The \textit{alignment block} gave the LLM the statement that each \probVal{} would more closely align with from a different portion of the survey.
To illustrate this, the alignment statement for \personaVal{abiorange}{process-oriented} learners was:
\begin{quote}
    \footnotesize
    \textit{If I'm going to use a new feature or technology, I use very clear directions or help from someone else to learn it.}
\end{quote}
Whereas for \personaVal{timblue}{tinkering-oriented} learners, it was:
\begin{quote}
    \footnotesize
    \textit{In order to learn new technology, I tinker with it, constructing my own understanding of how it works.}
\end{quote}
All prompts, including the 11th control prompt, are listed in Appendix~\ref{app:prompts}.

\boldify{Each system prompt could be broken down into three three blocks---an intro and two question blocks.}

\begin{table*}[h]
\begin{minipage}[t]{0.32\linewidth}
    {\begin{center}
        \textbf{Fibonacci Sequence}
    \end{center}}\par
    \cobolCodeBlock{}{fib}{1}
\end{minipage}
\hfill
\begin{minipage}[t]{0.27\linewidth}
    \begin{center}
        \textbf{Infinite Loop}
    \end{center}\par
    \cobolCodeBlock{}{inf_loop}{1}
\end{minipage}
\hfill
\begin{minipage}[t]{0.4\linewidth}
    \begin{center}
        \textbf{Rock-Paper-Scissors}
    \end{center}\par
    \cobolCodeBlock{}{rps}{1}
\end{minipage}
    \caption{The three COBOL programs used in our work.
    To see whether LLMs could discern the code, we removed superficial cues (e.g., variable names, program names).
    In total, we generated 1,072 explanations from these three programs.}
    \label{tab:COBOL_progs}
\end{table*}


\paragraph{Selecting COBOL Programs to Explain}
\label{subsubsec:cobol}

\boldify{Here's how we encouraged pro software engineers to engage with LLM code-explanations in a research setting.}

We chose to ground our investigation in code explanations for the COBOL programming language.
We chose COBOL because of its continued critical role in financial and government sectors~\cite{lee2020cnn}, with estimates suggesting that 95\% of ATM transactions (and 80\% of in-person banking activities) rely on COBOL~\cite{mentonelli2025cobol}.
Furthermore, COBOL is a legacy programming language, where the demand for developers with relevant COBOL expertise exceeds the supply ~\cite{martin2020brush}.
Thus, COBOL program explanations are ecologically valid in modern contexts.

\boldify{The programs needed to be at the right level of difficulty, challenging enough to explain but interesting enough to pique interest}

We selected COBOL programs based on three criteria.
The first was to select programs that were self-contained, excluding programs with external dependencies or data sources.
The second was to select programs that were not too simple, where explanations would be unlikely to require adaptation, consistent with Bunt et al.'s observations on the need for explanation~\cite{bunt2012explanations}.
The last was to select programs that were not too complex, where the LLMs' explanations may have become unsound through hallucination~\cite{kulesza2015principles}.

\boldify{Thus, we curated a set of 9 COBOL programs, asked 5 software engineers to help us narrow it down, and then obfuscated}

We curated a set of eight COBOL programs from the open-source accompaniment of \citet{coughlan2014beginning}'s introductory COBOL course\footnote{https://github.com/Apress/beg-cobol-for-programmers/tree/master}.
We supplemented this set with a Fibonacci number generator\footnote{http://progopedia.com/example/fibonacci/341/}, given its popularity in first-year CS courses.
In a group decision-making exercise, five professional software engineers from IBM collaboratively reviewed and eliminated six programs that were either too easily understood, to difficult to understand, or did not possess enough technical depth.

We obfuscated the remaining three programs to remove superficial cues that could make their purpose easily inferrable by the LLMs.
Specifically, we renamed program identifiers to \programmingfont{``MyProgram,''} renamed variables, renamed program and method identifiers, and removed code comments.
Table~\ref{tab:COBOL_progs} shows the three COBOL programs: 
the Fibonacci sequence generator, a program containing an infinite loop (to see if LLMs could detect the logic bug), and a rock-paper-scissors simulator%
\footnote{Our obfuscation efforts were successful.
The LLMs never identified the Rock-Paper-Scissors game or found the logic bug.}.

\paragraph{LLM Selection \& Explanation Generation}
\label{subsubsec:LLMs}

\boldify{Here's how we selected which model to use in our experiment.}

We conducted this experiment between June and September 2024, selecting open-weight LLMs (to promote replicability) that were approved for use on our institution's AI inference platform (IBM's Big AI Models).
Six models fit these criteria:
\begin{itemize}
    \item \llamaThreeIntro{}
    \item \codeLlamaIntro{}
    \item \ibmLlamaQIntro{}
    \item \graniteChatIntro{}
    \item \graniteCodeInstructIntro{}    
    \item \mixtralInstructIntro{}  
\end{itemize}

We performed zero-shot prompting \cite{kojima2022large} on these LLMs using the 11 prompts we developed.
We chose zero-shot prompting to reduce development overhead and preserve adaptation validity.

We generated 1,072 code explanations across all prompts, and we set only three parameters.
Two of the parameters governed the number of new tokens generated, which we set to be between 10--1,024 tokens.
Our reasoning was that this wide range would enable LLMs to produce both succinct and verbose explanations. 
The last parameter we set was the temperature, setting it to 0.5 to balance output consistency with response variability%
\footnote{As $t \rightarrow 0$, models become more deterministic across independent trials.
As $t \rightarrow 1$, models become more stochastic.}. 
We held these parameters constant throughout our investigation.

\paragraph{Analysis Methods}
\label{subsec:ngram-analysis}

Our investigation focuses solely on whether LLMs made linguistic adaptations in code explanations (rather than explanation accuracy), so we first must define what we mean by ``adaptation''.
\begin{definition*}[Adaptation]
Let \textit{P} be a linguistic phrase of any length.
We qualify \textit{P} as an adaptation for a \probVal{} if either:
\begin{enumerate}
    \item \textit{P} was significantly more likely to occur in one \probVal{}'s explanations than it was in the opposing value \textbf{and} in the control.
    \item \textit{P} occurred \textbf{exclusively} in one \probVal{} and appeared in \textbf{at least} 10\% of those explanations. 
\end{enumerate}
\end{definition*}

After cleaning the explanations (e.g., lowercasing, removing punctuation), we used scikit-learn's CountVectorizer to produce n-grams ($n \in [3,15]$).
To remove redundant n-grams, we established two exclusion criteria:
1) the tokens in the n-gram had to appear in the same order in an (n+1)-gram
2) the number of n-gram instances was equal to the number of (n+1)-gram instances.
This process left 442,216 n-grams.


\boldify{To treat all of these equally in an explanation, we programatically tagged these with $<$division\_name$>$ and counted instances}

\boldify{From each count, we could calculate how much more/less likely one phrase was to occur, using proportions and odds.}

\boldify{Given the quantity of n-grams, we wanted to give ourselves more confidence, so we created three thresholds to look at these data through.}

To satisfy both adaptation definition conditions, we computed the odds of each n-gram occurring in each \probVal{}, forming a distribution of odds.
To meet our first condition, we excluded n-grams where the odds fell below the midpoint of the distribution, leaving ``more likely than not'' n-grams.
For the second condition, we looked at all n-grams that occurred in \textit{only one} \probVal{} and excluded the n-grams that did not appear in at least 10\% of our data.
This yielded 1,029 n-grams.

Three authors collaboratively clustered the n-grams by conceptual similarity, which were independently validated by two non-author researchers.
Disagreements were resolved during collaborative discussions by changing a n-gram’s cluster or creating a new cluster until consensus was reached.
Then, regular expressions were made for each category.
These regular expressions were applied to all code explanations, counting the presence of each category.

\boldify{From these data, three authors clustered the n-grams via conventional content analysis, and once they reached data saturation, the larger team validated the clusters.}


\begin{table}[h]
    \small
    \centering
    \begin{tabularx}{\columnwidth}
    {@{}rp{0.475\linewidth}ccccc}
    \toprule
    \multicolumn{2}{l}{\textbf{Adaptation Category}} &
    \textbf{L} &
    \textbf{S} &
    \textbf{R} &
    \textbf{M} &
    \textbf{I} \\
    \midrule
    1. & \explainStepByStep{} & 
    \categoryTableCell{abiorange}{\uparrow} & 
    \categoryTableCell{abiorange}{\uparrow} & 
    & 
    & 
    \\ 
    2. & \questionsClarifications{} & 
    \categoryTableCell{timblue}{\uparrow}  & 
    \categoryTableCell{abiorange}{\uparrow}  & 
    & 
    &
    \categoryTableCell{timblue}{\uparrow}  
    \\ 
    3. &\inferUserInterests{} &
    \categoryTableCell{timblue}{\uparrow} & 
    & 
    \categoryTableCell{timblue}{\uparrow} & 
    \categoryTableCell{timblue}{\uparrow} &
    \\ 
    4. &\userOpinion{} &
    \categoryTableCell{timblue}{\uparrow} & 
    & 
    & 
    \\ 
    
    \midrule
    5. &\normalizeKnowledgeGaps{} & 
    & 
    \categoryTableCell{abiorange}{\uparrow} & 
    & 
    & 
    \\ 
    6. &\expressHelpfulness{} &
    & 
    \categoryTableCell{abiorange}{\uparrow} & 
    & 
    &
    \categoryTableCell{abiorange}{\uparrow} 
    \\ 
    7. &\hedgingLanguage{} &
    & 
    \categoryTableCell{abiorange}{\downarrow} & 
    & 
    \categoryTableCell{timblue}{\downarrow} &
    \\ 
    \midrule
    8. &\provideAssurance{} & 
    &
    & 
    \categoryTableCell{abiorange}{\uparrow} & 
    & 
    \\ 
    9. &\identifyRisks{} &
    & 
    & 
    \categoryTableCell{abiorange}{\uparrow}& 
    \\ 
    10. &\codeClarityComment{}&
    & 
    & 
    \categoryTableCell{abiorange}{\uparrow} & 
    \\
    \midrule
    11. & \codePurposeOverview{} &
    \categoryTableCell{timblue}{\downarrow} & 
     & 
    & 
   \categoryTableCell{timblue}{\downarrow} &   \\
    
    \midrule
    12. & \conciseExpl{} &
    & 
    & 
    & 
    &
    \categoryTableCell{timblue}{\uparrow}
    \\ 
    13. &\provideCodeSummary{} &
    & 
    & 
    & 
    &
    \categoryTableCell{abiorange}{\uparrow}
    \\ 

    \midrule
    &
    \multicolumn{1}{r}{\textbf{Totals:}} &
    5 & 
    5 & 
    4 & 
    3 &
    4 \\
    \bottomrule
    \end{tabularx}

    \caption{The 13 adaptations (rows).
    Each cell's color shows the \probVal{} (e.g., \personaVal{abiorange}{process-oriented}) the LLMs increased ($\uparrow$) or decreased ($\downarrow$) the frequency for, relative to its opposing value (e.g., \personaVal{timblue}{tinkering-oriented}) and the control.
    L: Learning Style, S: Self-Efficacy, R: Risk Attitudes, M: Motivations, I: Information Processing Style.}
    \label{tab:adaptation_summary_table}

\end{table}


\section{Results  
\draftStatus{**}
}
\label{sec:llm_adaptation_results}

\paragraph{How Did These LLMs Adapt?}

\boldify{RQ1 asks (a) \& (b). The answer to both is in  Table~\ref{tab:adaptation_summary_table} via 13 adaptations.}

We start with an overview answer for our research question.
Table~\ref{tab:adaptation_summary_table} shows a taxonomy of 13 linguistic adaptations (rows) that LLMs made in their code explanations, along with their associated \probType{s} (columns L--I).
Each cell reflects whether that adaptation was more ($\uparrow$) or less ($\downarrow$) likely to occur, and the cell colors reflect the adaptations' associated \probVal{}.
For instance, a \personaVal{abiorange}{$\uparrow$} in the L column represents an increased likelihood for \personaVal{abiorange}{process-oriented} learning style.
Similarly, a \personaVal{timblue}{$\downarrow$} in the M column represents a decreased likelihood for \personaVal{timblue}{tech-oriented} motivations.

The LLMs achieved generally good coverage for these 10 \probVal{s}. 
They had at least one adaptation in 8 out of the 10 values, with only \personaVal{timblue}{higher} self-efficacy and \personaVal{abiorange}{task-oriented} motivations adaptations missing.
This points to the efficacy of the 10 prompts in generating \probVal{} adaptations.

\boldify{Now that you've been primed, here's the takeaway}

\paragraph{Learning Style Adaptations
\draftStatus{AAA}{2.5}}
\label{subsec:learning_style_adapt}

\begin{table*}
    \small
    \centering
    \begin{NiceTabular}
    {X@{}r|ccc|ccc|ccc|cccX}
    \toprule
        &
        \multirow[b]{2}{*}{\textbf{Adaptation Category}} & 
        \textit{Unad.} & 
        \textit{\personaVal{abiorange}{Proc.}} & 
        \textit{\personaVal{timblue}{Tink.}} & 
        \multicolumn{3}{c|}{\textit{\personaVal{abiorange}{Proc.} vs. Unad.}} & \multicolumn{3}{c|}{\textit{\personaVal{timblue}{Tink.} vs. Unad.}} & \multicolumn{3}{c}{\textit{\personaVal{abiorange}{Proc.} vs. \personaVal{timblue}{Tink.}}} & \\ & & 
        $\hat{\pi}_u$ & 
        $\hat{\pi}_p$ &  
        $\hat{\pi}_t$ & 
        $\hat{\phi}$ & 
        $p$ & 
        $d$ & 
        $\hat{\phi}$ & 
        $p$ & 
        $d$ & 
        $\hat{\phi}$ & 
        $p$ & 
        $d$ & \\
        \midrule
        &
        \explainStepByStep{} & 
        .30& 
        \textbf{.56} & 
        .26 & 
        \cellcolor{abiorange}{2.97 $\uparrow$} & 
        \cellcolor{abiorange}$<$ .001 & 
        \cellcolor{abiorange}0.60 & 
        
        $\approx$ 1.00 & 
        .365 & 
        0.09 & 
        
        \cellcolor{abiorange}{3.49 $\uparrow$} & 
        \cellcolor{abiorange}$<$ .001 & 
        \cellcolor{abiorange}0.69 
        & \\
        &
        \questionsClarifications{} & 
        .10 & 
        .16 & 
        \textbf{.27} & 
        
        1.73 $\uparrow$ & 
        .140 & 
        0.30 &
        
        \cellcolor{timblue}{3.30 $\uparrow$} & 
        \cellcolor{timblue}.002 & 
        \cellcolor{timblue} 0.66 & 
        
        \cellcolor{timblue}{1.97 $\uparrow$} & 
        \cellcolor{timblue}.049 & 
        \cellcolor{timblue}0.35 &\\&
        \inferUserInterests{} & 
        .00& 
        .00& 
        \textbf{.22} & 
        \multicolumn{3}{c|}{---} & 
        \multicolumn{3}{c|}{---} & 
        \multicolumn{3}{c}{---} &\\&
        \userOpinion{} & 
        .00& 
        .00& 
        \textbf{.11} & 
        \multicolumn{3}{c|}{---} & 
        \multicolumn{3}{c|}{---} & 
        \multicolumn{3}{c}{---} &\\
        \bottomrule
    \end{NiceTabular}
    \caption{Statistical details of the first four adaptations. 
    \textit{Unad.}: unadapted explanations. \personaVal{abiorange}{Proc.}: explanations \personaVal{abiorange}{process-oriented} adaptations. \personaVal{timblue}{Tink.}:  \personaVal{timblue}{tinkering-oriented} adaptations.
    \textbf{Cols 2--4:} The proportions of unadapted ($\hat{\pi}_u$), process-oriented ($\hat{\pi}_p$) adapted, and tinkering-oriented ($\hat{\pi}_t$) adapted explanations with each adaptation.
    \textbf{Cols 5--13:} The three odds ratio (OR) comparisons, showing how much more ($\uparrow$) or less ($\downarrow$) likely an adaptation was, its significance ($p$), and effect size (Cohen's $d$).
    Dashes: undefined comparison (divide by 0).}
    \label{tab:01_learning_style_adaptations}
\end{table*}

\boldify{The LLMs adapted in four ways for different learning styles: (1) providing Step-by-Step Explanations (process), (2) inviting Questions/Clarifications (tinker), (3) Asking the users's thoughts (tinker), and (4) suggesting things of possible Interest (tinker).}

Table~\ref{tab:01_learning_style_adaptations} (rows) details the first four adaptations in Table~\ref{tab:adaptation_summary_table}.
For each row, the first three columns show the proportions of unadapted ($\hat{\pi}_u$), \personaVal{abiorange}{process-oriented} adapted ($\hat{\pi}_p$), and \personaVal{timblue}{tinkering-oriented} adapted ($\hat{\pi}_t$) explanations with that adaptation.
For example, 30\% of the unadapted explanations had \explainStepByStep{} instances, compared to 56\% of the \personaVal{white}{process-oriented} adapted explanations.
These presence/absence outcomes naturally lent themselves to odds ratio analyses%
\footnote{Appendix~\ref{app:prop_odd_OR} provides a short primer on proportions ($\hat{\pi}$), odds ($\hat{\omega}$), and odds ratios ($\hat{\phi}$).}.

In the \explainStepByStep{} adaptation, the LLM stated that it would structure its explanation in methodical ways, like in numeric steps or by division:

\llmAdaptQuote{\mixtralInstruct{}}
{abiorange}
{process-oriented}
{Here's a step-by-step explanation of the process:\newline
1. The program starts at the Begin section, where the program...\newline
2. Counter C1 starts at 13 and decreases by 5 each time...
}
{Fib-8/1/24}
{Here's a step-by-step explanation of the process:
1. The program starts at the Begin section, where the PERFORM statement initializes and varies the counters.
2. Counter C1 starts at 13 and decreases by 5 each time until it reaches a value less than 2.
3. For each value of C1, counter C2 starts at 15 and decreases by 4 each time until it reaches a value less than 1.
4. For each pair of C1 and C2 values, counter C3 starts at 1 and increases by 1 each time until it reaches a value greater than 5.
5. The DISPLAY statement in the Loop section shows the current values of the three counters in each iteration.
6. The loop repeats until all counters reach their specified bounds, and then the program stops.}
\llmAdaptQuote{\llamaThree{}}
{abiorange}
{process-oriented}
{Let's break down the code step by step:
\newline
**DATA DIVISION** This section defines the variables...
\newline
**PROCEDURE DIVISION** This section contains...
}
{Fib-8/1/24}
{Let's break down the code step by step:
\newline
**Data Division**: This section defines the variables used in the program. We have seven variables:
\newline
* `F1`, `F2`, and `F3` are numeric variables that can hold three-digit numbers (pic 999).
* `I` is a numeric variable that can hold two-digit numbers (pic 99).
* `FST` is a character variable that can hold three characters (pic XXX).
* `RES` is a character variable that can hold a string of up to 64 characters (pic X(64)).
\newline
**Procedure Division**: This section contains the program logic.\newline
1. The program initializes the variables: `I` is set to 0, `F1` and `F2` are set to 0 and 1, respectively, and `RES` is set to an empty string.
2. The program enters a loop that will run until `I` is greater than 15.}

Table~\ref{tab:01_learning_style_adaptations}'s remaining nine columns show the three odds ratio (OR) statistical tests.
Each test shows how much more likely ($\uparrow$) these adaptations were to occur, including significance ($p$) and effect size ($d$)%
\footnote{Haddock et al. (\citeyear{haddock1998using}) found that $\ln(\hat{\phi})\left(\sqrt{3/\pi}\right) \approx d$.}.
For instance, the \textbf{\explainStepByStep{}} adaptation was almost 3 times more likely (2.97$\uparrow$) to occur in \personaVal{abiorange}{process-oriented} explanations than in the unadapted explanations.
The adaptation was even more likely (3.49$\uparrow$) to occur in \personaVal{white}{process-oriented} explanations than in \personaVal{timblue}{tinkering-oriented} ones.
Both likelihood differences were statistically significant ($p < .001$) with medium effect sizes, with Cohen's $d \in [0.5,0.8)$%
\footnote{We consider Cohen's $d < .02$ no effect, $d \in [0.2, 0.5)$ a small effect, $d \in [0.5, 0.8)$ a medium effect, and $d \geq 0.8$ a large effect~\cite{cohen2013statistical}.}.



\boldify{The LLMs directed the first adaptation, \explainStepByStep{}, toward process-oriented users---those who like to start with a process, then learn the details in a "top-down" direction. In this adaptation...}

\boldify{how do we know these were for process-oriented learners? 2 reasons: 1st, they told us...}

Two additional sources of evidence triangulate why the \explainStepByStep{} adaptation was a \personaVal{abiorange}{process-oriented} learning style adaptation, the first source being the LLMs themselves.
Almost 20\% of explanations with this adaptation stated that they took this methodical approach \textit{because} of the user's learning style:

\llmAdaptQuote{\mixtralInstruct{}}
{abiorange}
{process-oriented}
{[The explanation is] tailored for a process-oriented learner, with clear instructions and step by step explanations.}
{8/1/24-RPS}
{}


\boldify{**...2nd reason, odds said process-oriented more likely, whereas unadapted and tinkering less likely}

\boldify{Research around learning styles suggests why this makes sense.}

The second evidence source comes from prior GenderMag literature.
While using GenderMag's walkthrough, \citet{chatterjee2024debugging}'s most frequent \personaVal{abiorange}{process-oriented} issue was ``a lack of step-by-step guidance'', which blocked such learners from making progress in online courseware.
This points to why the \explainStepByStep{} adaptation was appropriate for process-oriented learners.


\boldify{LLMs' directed their other 3 adaptations toward tinkering-oriented learners---those who prefer to explore details first in a "bottom-up" direction. 
One such adaptation was the \questionsClarifications{}... }


In Table~\ref{tab:01_learning_style_adaptations}'s second adaptation, \textbf{\questionsClarifications{}}, the LLMs invited further engagement from the user:

\llmAdaptQuote{\ibmLlamaQ{}}
{timblue}
{tinkering-oriented}
{Do you have any specific questions about the program...}
{8/8/24-RPS}
{do you have any specific questions about the program or its functionality}

\llmAdaptQuote{\codeLlama{}}
{timblue}
{tinkering-oriented}
{If you have any further questions, feel free to ask.}
{8/6/24-Fib}
{}


\boldify{How  we classify these as tinkering oriented: stats tell us so.}

The data showed that the \questionsClarifications{} adaptation (Table~\ref{tab:01_learning_style_adaptations}) was geared towards \personaVal{timblue}{tinkering-oriented} learners;
it was almost twice as likely to occur in tinkering-oriented explanations than in \personaVal{abiorange}{process-oriented} ones ($1.97\uparrow$), and more than three times more likely than in unadapted ($3.30\uparrow$). 
Both differences were statistically significant ($p<.05$) with small-to-medium effect sizes.

\boldify{Unlike the \explainStepByStep{}, none of these instances explicitly tied the \probVal{} as the reason why, but they still left clues; observe its experimentation language.}

Over on third of the \questionsClarifications{} instances (38.5\%) revealed this adaptation's connection to the \personaVal{timblue}{tinkering-oriented} learning style.
These \questionsClarifications{} instances invited the tinkering-oriented users to engage, asking if they wanted to know how the code worked or how to experiment with it:
\llmAdaptQuote{\llamaThree{}}
{timblue}
{tinkering-oriented}
{Do you have any questions about \textbf{how [the code] works} or how you could modify it?}
{9/3/24-RPS}
{do you have any questions about how it works or how you could modify it}
\llmAdaptQuote{\llamaThree{}}
{timblue}
{tinkering-oriented}
{Do you have any questions about how [the code] works, or would you like to \textbf{experiment with} modifying it to see how it behaves?}
{9/4/24-RPS}
{do you have any questions about how it works or would you like to experiment with modifying it to see how it behaves}
This language mirrored the language in two statements in the \personaVal{timblue}{tinkering-oriented} prompt, connecting this adaptation to this learning style:
\begin{quote}
    \footnotesize \textit{In order to learn new technology, I tinker with it, constructing my own understanding of \textbf{how it works}.}

    \textit{My first step in learning new technology is \textbf{experimenting} and tinkering with it.}
\end{quote}

%
%

\boldify{These questions encourage a level of reflection}

The STEM education literature provides another source of evidence to explain why the \questionsClarifications{} adaptation was for \personaVal{timblue}{tinkering-oriented} learners.
In particular, \citet{vossoughi2013tinkering} suggest that offering suggestions or learning about someone's ideas, questions, and goals to support the development of engineering projects lies at the heart of tinkering, a position supported by \citet{petrich2013looks}.
The \questionsClarifications{} adaptation appears to lean into that pedagogical approach to learn about tinkering-oriented learners' questions or goals.

\boldify{The \userOpinion{} adaptation was different, occurring exclusively for one \probVal{}, and it was an opposite of the \questionsClarifications{}}

\boldify{The second category which appeared in only one \probVal{} was the \inferUserInterests{}, and Figure~\ref{fig:results_learn_04_usr_cmmtry} shows its prevalence}

\boldify{Table~\ref{tab:learn_04_usr_cmmtry_all_statements} shows the exhaustive list of observations, clustered by type. }

The last two adaptations from Table~\ref{tab:01_learning_style_adaptations} illustrate the second kind of adaptation from Section~\ref{sec:methods}, occurring \textit{exclusively} in \personaVal{timblue}{tinkering-oriented} explanations.
The \inferUserInterests{} occurred in 22.1\% of tinkering-oriented explanations, where the model said something relating to the user's interests.
The most frequent example of these occurred in the opening sentence of 85.7\% explanations, where the LLM stated that the tinkering-oriented user was a COBOL enthusiast:
\llmAdaptQuote{\llamaThree{}}
{timblue}
{tinkering-oriented}
{A COBOL enthusiast, eh?}
{7/31/24-Fib}
{}

\boldify{This comment was unexpected; it had no ties to the prompt, and it has no tie to the literature.}

Despite its prevalence, this adaptation was a surprise for two reasons.
First, neither the \personaVal{timblue}{tinkering-oriented} prompt nor the existing learning style literature supported a relationship between \personaVal{timblue}{tinkering-oriented} learning style and COBOL enthusiasm.
Second, every instance of this adaptation came from \llamaThree{}.
We remain open to the possibility that \llamaThree{}'s training data may have found this relationship.

\boldify{The next most-frequent comment that \llamaThree{} made was on its user's propensity for exploration.}



The second adaptation that occurred exclusively for \personaVal{timblue}{tinkering-oriented} learners was the \textbf{\userOpinion{}} adaptation, where the LLMs asked an open-ended, reflective question to the user:

\llmAdaptQuote{\llamaThree{}}
{timblue}
{tinkering-oriented}
{Now I'm curious, what do you think about this program?}
{8/8/24-RPS}
{}
\llmAdaptQuote{\llamaThree{}}
{timblue}
{tinkering-oriented}
{...what do you think this program is intended to do?}
{8/11/24-RPS}
{}


\begin{table*}[h]
    \small
    \centering
    \begin{NiceTabular}
    {X@{}r|ccc|ccc|ccc|ccc}
    \toprule
    &
        \multirow[b]{2}{*}{\textbf{Adaptation Category}}
        & \textit{Unad.}
        & \textit{\personaVal{abiorange}{Low}}
        & \textit{\personaVal{timblue}{High}}
        & \multicolumn{3}{c|}{\textit{\personaVal{abiorange}{Low} vs. Unad.}}
        & \multicolumn{3}{c|}{\textit{\personaVal{timblue}{High} vs. Unad.}} 
        & \multicolumn{3}{c}{\textit{\personaVal{abiorange}{Low} vs. \personaVal{timblue}{High}}} \\ &
        & $\hat{\pi}_u$ & 
        $\hat{\pi}_l$ & 
        $\hat{\pi}_h$ &
        $\hat{\phi}$ & 
        $p$ & 
        $d$ & 
        $\hat{\phi}$ & 
        $p$ &
        $d$ & 
        $\hat{\phi}$ & 
        $p$ & 
        $d$ \\
        \midrule
        &
        \normalizeKnowledgeGaps{} & 
        .00 &  
        \textbf{.28} & 
        .01 & 
        \multicolumn{3}{c|}{---} & 
        \multicolumn{3}{c|}{---} &
        
        \cellcolor{abiorange}{35.67 $\uparrow$} & 
        \cellcolor{abiorange}$<$ .001 & 
        \cellcolor{abiorange}1.97 \\&
        
        \expressHelpfulness{} & 
        .13 & 
        \textbf{.47} & 
        .24 & 
        
        \cellcolor{abiorange}{5.73 $\uparrow$} & 
        \cellcolor{abiorange}$<.001$ &
        \cellcolor{abiorange}0.96 &
        
        \cellcolor{timblue}{2.05 $\uparrow$} & 
        \cellcolor{timblue}.041 & 
        \cellcolor{timblue}0.40 & 
        
        \cellcolor{abiorange}{2.79 $\uparrow$} & 
        \cellcolor{abiorange}$<.001$ & 
        \cellcolor{abiorange}0.57 \\&
        
        \hedgingLanguage{} & 
        .16 & 
        \textbf{.01} &  
        .15 & 
        \cellcolor{abiorange}18.73 $\downarrow$ & 
        \cellcolor{abiorange}$<.001$  & 
        \cellcolor{abiorange}1.61 &
        $\approx$ 1 & 
        .446 & 
        0.07 &
        \cellcolor{abiorange}16.38 $\downarrow$ & 
        \cellcolor{abiorange}$<.001$  & 
        \cellcolor{abiorange}1.54 \\

        \bottomrule
    \end{NiceTabular}
    \caption{Statistical summary of the self-efficacy adaptations, in the same format as Table~\ref{tab:01_learning_style_adaptations}. 
    \textit{Unad.}: unadapted explanations. \personaVal{abiorange}{Low}: \personaVal{abiorange}{lower} self-efficacy. \personaVal{timblue}{High}: \personaVal{timblue}{higher} self-efficacy.
    The \normalizeKnowledgeGaps{} adaptation was the strongest adaptation, with most instances normalizing a gap in COBOL knowledge.
    }
    \label{tab:02_self_efficacy_adaptations}
\end{table*}


\boldify{Tinkering literature has pointed to a problem --- that tinkerers can tinker to excess, and some LLMs even stated this in their explanations!}

Prior tinkering-related works suggest why the \userOpinion{} adaptation was a \personaVal{timblue}{tinkering-oriented} one.
In particular, the last of \citeauthor{dron2014ten}'s (\citeyear{dron2014ten}) \textit{``10 principles for effective tinkering''} calls for tinkerers to reflect while tinkering, telling stories about their reflection.
The open-ended structure of the \userOpinion{} adaptation, which rely on answers beyond binary responses, may be one such method to encourage reflective tinkering.


\paragraph{Self-Efficacy Adaptations%
\draftStatus{AAA}{2.5}
}
\label{subsec:self_efficacy_adapt}

\boldify{Now on to diverse self-efficacies, and here's what this means.}


Table~\ref{tab:02_self_efficacy_adaptations} shows the adaptations from rows 5--7 in Table~\ref{tab:adaptation_summary_table}.
This table shows the proportions of unadapted ($\hat{\pi}_u$), \personaVal{abiorange}{lower} self-efficacy ($\hat{\pi}_l$), and \personaVal{timblue}{higher} self-efficacy ($\hat{\pi}_h$) explanations.


\boldify{The first of these categories manifested by providing supportive language to the user, and here's what it looked like}

The first of these adaptations was \textbf{\normalizeKnowledgeGaps{}}.
In this adaptation, the LLM addressed any potential knowledge gaps, with most instances (85.7\%) normalizing the user's knowledge gap in COBOL:


\llmAdaptQuote{\llamaThree{}}
{abiorange}
{lower}
{Don't worry if you're not familiar with COBOL.}
{8/30/24-Fib}
{}


\boldify{As Table~\ref{tab:02_self_efficacy_adaptations} shows, this adaptation overwhelmingly favored lower self-efficacy }

In these data, the \normalizeKnowledgeGaps{} adaptation had the largest likelihood difference between two treatments.
It was 35 times more likely to occur in \personaVal{abiorange}{lower} self-efficacy explanations than in \personaVal{timblue}{higher} ones, and it \textit{never} appeared in the unadapted explanations.
It was statistically significant, with the largest effect size ($d:1.97$).

\boldify{For lower self-efficacy, such instances were almost evenly split --- about half just said ``don't worry if you're not familiar with COBOL,'' but the other half said \textit{why} }

\normalizeKnowledgeGaps{} instances came in three flavors.
The first stopped at normalizing the knowledge gap:
\llmAdaptQuote{\llamaThree{}}
{abiorange}
{lower}
{Don't worry if you're not familiar with the Fibonacci sequence}
{7/31/24-Fib}
{don't worry if you're not familiar with the fibonacci sequence}
The second flavor also explained \textit{why} a knowledge gap may exist.
For instance, the LLMs sometimes pointed to COBOL's age:
\llmAdaptQuote{\llamaThree{}}
{abiorange}
{lower}
{\textcolor{gray}{Don't worry...familiar with COBOL -} it's an older programming language and it can be a bit tricky...}
{8/8/24-RPS}
{don't worry if you're not familiar with cobol it's an older programming language and it can be a bit tricky to understand at first}

\llmAdaptQuote{\llamaThree{}}
{abiorange}
{lower}
{\textcolor{gray}{Don't worry...familiar with cobol -} it's an older programming language and it can be a bit challenging...}
{8/24/24-RPS}
{don't worry if you're not familiar with cobol it's an older programming language and it can be a bit challenging to read}

With the last flavor, the LLMs told the user that they were \textit{a solution} to that knowledge gap, that they would help:
\llmAdaptQuote{\llamaThree{}}
{abiorange}
{lower}
{\textcolor{gray}{...not familiar with cobol} - I'm here to help you understand what this code does.}
{8/8/24-Fib}
{Don't worry if you're not familiar with cobol - I'm here to help you understand what this code does.}

\llmAdaptQuote{\llamaThree{}}
{abiorange}
{lower}
{\textcolor{gray}{...this code looks unfamiliar}  - I'm here to help you understand it}
{8/31/24-RPS}
{Don't worry if this code looks unfamiliar i'm here to help you understand it}

\boldify{Normalizing knowledge gaps has been tied to psychological safety, which is something that }

Normalizing knowledge gaps can promote psychological safety, which helps people learn and perform better at their tasks~\cite{Edmondson1999PsychSafety}.
Additionally \citet{Jones2024PsychSafety} found a positive correlation between psychological safety and help-seeking, which \citet{huet2016motivation} found that \personaVal{abiorange}{lower} self-efficacy people were more likely to do.
As such, these speak to the appropriateness of the \normalizeKnowledgeGaps{} adaptation's prevalence in lower self-efficacy explanations.




\boldify{Here is the next adaptation category, the \expressHelpfulness{}}



The second adaptation from Table~\ref{tab:02_self_efficacy_adaptations} was to \expressHelpfulness{}.
In this adaptation, the LLM expressed that it hoped the explanation had served its  the user:



\llmAdaptQuote{\ibmLlamaQ{}}
{abiorange}
{lower}
{I hope this helps you understand what the code does.}
{8/1/24-Fib}
{i hope this helps you understand what the code does}

\llmAdaptQuote{\mixtralInstruct{}}
{abiorange}
{lower}
{I hope this explanation helps you understand the code.}
{7/31/24-Fib}
{i hope this explanation helps you understand the code better}

\boldify{As with before, this adaptation category was a complementary adaptation for self-efficacy, and the first step is showing that it is tied to lower self-efficacy}

The \textbf{\expressHelpfulness{}} adaptation was the first linguistic adaptation that adapted for one of the problem-solving style \textit{types}.
Illustrating this, the adaptation was significantly more likely to occur in both \personaVal{abiorange}{lower} self-efficacy (\personaVal{abiorange}{5.73$\uparrow$}) \textit{and} \personaVal{timblue}{higher} self-efficacy explanations (\personaVal{timblue}{2.05$\uparrow$}) than they were in unadapted ones.
However, we still classify this as a \personaVal{abiorange}{lower} self-efficacy adaptation because it was significantly more likely to occur in lower than in higher self-efficacy explanations (2.79$\uparrow$), with a medium effect size.

\boldify{The self-efficacy prompt may point to why the LLMs used these two adaptations}

The \expressHelpfulness{} adaptation's prevalence at the self-efficacy \probType{} level may have come from the statements in the self-efficacy prompt.
In particular, three statements pertained to accessing help:

\begin{quote}
    \footnotesize
    I am able to use unfamiliar technology when...
    \begin{itemize}[ label = {},itemsep=2pt, parsep = 0pt, topsep=0pt]
        \item ...I can call someone for help if I get stuck
        \item ...no one is around to help if I need it.
        \item ...I have just the built-in help for assistance.
    \end{itemize}
\end{quote}
One possibility is that for the LLMs, it did not matter whether someone disagreed (\personaVal{abiorange}{lower}) or agreed (\personaVal{timblue}{higher}) with these statements;
they still wanted to make sure that everyone received the help they needed.

\boldify{Here's the \hedgingLanguage{} category, and here's what it looked like.}

All adaptations so far have \textit{increased} linguistic frequency, but the \textbf{\hedgingLanguage{}} was the first to \textit{decrease} this frequency.
In our data, \hedgingLanguage{}{} occurred when the LLMs hedged their explanations, instead of using definitive language:


\llmAdaptQuote{\graniteCodeInstruct{}}
{abiorange}
{lower}
{Overall, this cobol code \textbf{appears to be} a simple program...}
{8/8/24-RPS}
{overall this cobol code appears to be a simple program that takes input from the user performs some evaluation based on certain conditions and displays the result}


\boldify{Figure~\ref{fig:results_se_07_hedge_lang} shows why this adaptation was different --- the LLMs drew back on using this hedging language for lower self-efficacy adapted code explanations }


As Table~\ref{tab:02_self_efficacy_adaptations} shows, the LLMs adapted how frequently they hedged their language in \personaVal{abiorange}{lower} self-efficacy explanations.
They were over 18 times \textit{less} likely to hedge their language in lower self-efficacy explanations than in unadapted ones (18.73$\downarrow$).
Similarly, they were over 16 times less likely to hedge their language in lower self-efficacy explanations than in \personaVal{timblue}{higher} self-efficacy ones (16.38$\downarrow$).
Both differences were statistically significant with large effect sizes.

\boldify{Hedging language has ties to confidence, so one possible explanation for why the LLMs drew back on this language is that they wanted to portray more confidence to those who might not have as much as others.}

Hedging language communicates confidence in a statement~\cite{hyland1996writing,coates1987epistemic}, and one explanation may be the LLMs adapting to the prompt.
The LLMs were told that \personaVal{abiorange}{lower} self-efficacy users were more likely to agree with:
\begin{quote}
    \footnotesize
    I am not confident about my ability to use and learn technology. I have other strengths.
\end{quote}
Prior works have shown that signaling uncertainty in LLM outputs can reduce over-reliance on those outputs, helping people feel more confident in their final answer~\cite{kim2024uncertaintyllm}.
Increased confidence can result in an increased likelihood that someone will persevere, a reported struggle for those with lower self-efficacy~\cite{bandura1994self}.


\paragraph{Risk Attitude Adaptations
\draftStatus{AAA}{2.5}}
\label{subsec:risk_adapt}

\begin{table*}
    \small
    \centering
    \begin{NiceTabular}
    {X@{}r|ccc|ccc|ccc|cccX}
    \toprule
    &
        \multirow[b]{2}{*}{\textbf{Adaptation Category}}
        & \textit{Unad.}
        & \textit{\cellcolor{abiorange}{Ave.}}
        & \textit{\cellcolor{timblue}{Tol.}}
        & \multicolumn{3}{c|}{\textit{\personaVal{abiorange}{Ave.} vs. Unad.}}
        & \multicolumn{3}{c|}{\textit{\personaVal{timblue}{Tol.} vs. Unad.}} 
        & \multicolumn{3}{c}{\textit{\personaVal{abiorange}{Ave.} vs. \personaVal{timblue}{Tol.}}} &\\&
        & $\hat{\pi}_u$ &
        $\hat{\pi}_a$ & 
        $\hat{\pi}_t$ &
        $\hat{\phi}$ & 
        $p$ & 
        $d$ & 
        $\hat{\phi}$ & 
        $p$ &
        $d$ & 
        $\hat{\phi}$ & 
        $p$ & 
        $d$ &\\
        \midrule
        &
        \provideAssurance{} & 
       .00 &  
        \textbf{.19} & 
       .00 & 
        \multicolumn{3}{c|}{---} & 
        \multicolumn{3}{c|}{---} &
        \multicolumn{3}{c}{---} &\\&
        
        \identifyRisks{} & 
       .00&
        \textbf{.22} &
        .01 &
        
        \multicolumn{3}{c|}{---} & 
        \multicolumn{3}{c|}{---} & 
        
        \cellcolor{abiorange}{27.68 $\uparrow$} & 
        \cellcolor{abiorange}$<.001$ & 
        \cellcolor{abiorange}1.83 &\\&
        
        \codeClarityComment{} & 
        .01 & 
        \textbf{.22} &  
        .11 & 
        \cellcolor{abiorange}{5.27 $\uparrow$} &
        \cellcolor{abiorange}$<.001$  & 
        \cellcolor{abiorange}0.92 &
        2.08 & 
        .146 & 
        0.404 &
        1.97 $\uparrow$ &  
        .076  & 
        0.40 &\\
        \bottomrule
    \end{NiceTabular}
    \caption{Statistical summary of the risk attitudes adaptations, in the same format as Table~\ref{tab:02_self_efficacy_adaptations}. 
    \textit{Unad.}: unadapted explanations. \personaVal{abiorange}{Ave.}: \personaVal{abiorange}{risk-averse}. \personaVal{timblue}{Tol.}: \personaVal{timblue}{risk-tolerant}.
    These adaptations targeted potential costs and risks for the user, including complexity.}
    \label{tab:03_risk_attitudes_adaptations}
\end{table*}


\begin{table*}
    \small
    \centering
    \begin{NiceTabular}
    {X@{}r|ccc|ccc|ccc|cccX}
    \toprule
    &
        \multirow[b]{2}{*}{\textbf{Adaptation Category}}
        & \textit{Unad.}
        & \textit{\cellcolor{abiorange}{Task}}
        & \textit{\cellcolor{timblue}{Tech}}
        & \multicolumn{3}{c|}{\textit{\personaVal{abiorange}{Task} vs. Unad.}}
        & \multicolumn{3}{c|}{\textit{\personaVal{timblue}{Tech} vs. Unad.}} 
        & \multicolumn{3}{c}{\textit{\personaVal{abiorange}{Task} vs. \personaVal{timblue}{Tech}}}& \\&
        & $\hat{\pi}_u$ &
        $\hat{\pi}_{ta}$ & 
        $\hat{\pi}_{te}$ & 
        $\hat{\phi}$ & 
        $p$ & 
        $d$ & 
        $\hat{\phi}$ & 
        $p$ &
        $d$ & 
        $\hat{\phi}$ & 
        $p$ & 
        $d$ &\\
        \midrule
        &
        \codePurposeOverview{} & 
        .57 &
        .54 &
        \textbf{.33} &
        
        $\approx$ 1 &
        .393 &
        0.07 &
        \cellcolor{timblue}{2.74 $\downarrow$} &
        \cellcolor{timblue}$<$ .001 &
        \cellcolor{timblue}0.55 &
        \cellcolor{timblue}{2.43 $\downarrow$} &
        \cellcolor{timblue}.002 &
        \cellcolor{timblue}0.49&\\

        \bottomrule
    \end{NiceTabular}
    \caption{Statistical summary of the motivations adaptations, in the same format as Table~\ref{tab:03_risk_attitudes_adaptations}. 
    \textit{Unad.}: unadapted explanations. \personaVal{abiorange}{Task}: \personaVal{abiorange}{task-oriented}. \personaVal{timblue}{Tech}: \personaVal{timblue}{tech-oriented}.
    This adaptation may have been the LLMs' attempt to keep the learning ``fun''.}
    \label{tab:05_mot_adaptations}
\end{table*}
\boldify{The third \probType{} from Table~\ref{tab:01-GenderMag-Persona-Table} was risk-attitudes, and here's what that means. The three adaptations }

Table~\ref{tab:03_risk_attitudes_adaptations} shows the three adaptations from rows 8--10 in Table~\ref{tab:adaptation_summary_table}.
This table shows the proportions of unadapted ($\hat{\pi}_u$), \personaVal{abiorange}{risk-averse} ($\hat{\pi}_a$), and \personaVal{timblue}{risk-tolerant} ($\hat{\pi}_t$) explanations.

\boldify{The first adaptation category was the \provideAssurance{}, and it appeared exclusively for the \personaVal{abiorange}{risk-averse} --- most of these combined both what the code \textit{was/did} and what the code \textit{was/did not}}

The first of these adaptations was \textbf{\provideAssurance{}}.
This adaptation occurred when the LLM provided assurances (or reassurances) to the user:
\llmAdaptQuote{\graniteChat{}}
{abiorange}
{risk-averse}
{However, I can assure you that...}
{8/1/24-Inf}
{however i can assure you that the provided code is correct and will function as intended when executed in a cobol compiler}

In these data, \textit{every} instance of the \provideAssurance{} adaptation occurred in \personaVal{abiorange}{risk-averse} explanations, providing compelling evidence that this adaptation was a risk-averse one.
Two additional sources of evidence also point to this conclusion.
The first source was that 95\% of the explanations with this adaptation mentioned that the assurances were because of the user's risk-averse attitude:
\llmAdaptQuote{\llamaThree{}}
{abiorange}
{risk-averse}
{Given your risk-averse attitude i want to assure you...}
{9/3/24-RPS}
{given your riskaverse attitude i want to assure you that this program is quite straightforward and doesn't involve any complex or risky operations}

The LLMs occasionally alluded to statements from the \personaVal{abiorange}{risk-averse} prompt.
For example, the LLMs were told that risk-averse people were more likely to agree with:
\begin{quote}
\footnotesize
    ``\textit{I avoid using new apps or technology before they are \textbf{well-tested}.}''
\end{quote}
10\% of these explanations mirrored this language while providing reassurances:
\llmAdaptQuote{\llamaThree{}}
{abiorange}
{risk-averse}
{...this code doesn't involve any complex or \textbf{untested features}}
{7/31/24-Inf}
{given your riskaverse nature i want to assure you that this code doesn't involve any complex or untested features}

\boldify{In these data, the LLMs also addressed the risks and complexity of the COBOL code outside of providing assurances --- still for the \personaVal{abiorange}{risk-averse}.}

The second adaptation from Table~\ref{tab:03_risk_attitudes_adaptations} was \textbf{\identifyRisks{}}, where the LLMs addressed whether the user would face any risks with the code:

\llmAdaptQuote{\mixtralInstruct{}}
{abiorange}
{risk-averse}
{The code is easy to understand and test, making it less likely to cause any unexpected issues or risks.}
{7/31/24-Fib}
{the code is easy to understand and test making it less likely to cause any unexpected issues or risks}

\llmAdaptQuote{\mixtralInstruct{}}
{abiorange}
{risk-averse}
{[The code] does not use any advanced features or take any risks with your data and it provides clear feedback based on the input you provide.}
{8/1/24-RPS}
{it does not use any advanced features or take any risks with your data and it provides clear feedback based on the input you provide}

The statistical evidence suggested that this was a \personaVal{abiorange}{risk-averse} adaptation.
Illustrating this, this adaptation was almost \textit{28 times} more likely to occur in risk-averse explanations than in \personaVal{timblue}{risk-tolerant} ones (\personaVal{abiorange}{27.68$\uparrow$}).
This likelihood difference was significant, with a large effect size.

Prior work have investigated how to generate LLM risk identification processes~\cite{Adejumo2025LLMExplainRisk}, and prior risk attitude literature helps explain the importance of the \provideAssurance{} and \identifyRisks{} adaptations for \personaVal{abiorange}{risk-averse} users.
When it comes to cost/benefit analyses, people with more risk-averse attitudes place a greater emphasis on the perceived costs (e.g.,~\parentheticalCite{Kind2017RiskAversion}; \parentheticalCite{Kramer2025RiskAversionCostBenefit}; \parentheticalCite{Li2026RiskAversionCostBenefit}), selecting safer actions when the outcomes are uncertain.
Proactive assurances of possible risks could become effective elucidation tools for uncertain outcomes, particularly when using unfamiliar technology.

The last adaptation in Table~\ref{tab:03_risk_attitudes_adaptations} was \textbf{\codeClarityComment{}}, where the LLM made observations about the code:

\llmAdaptQuote{\mixtralInstruct{}}
{abiorange}
{risk-averse}
{This is a simple, straightforward COBOL program...}
{8/1/24-Fib}
{This is a simple, straightforward COBOL program, without any advanced features or complex integrations.}

\llmAdaptQuote{\llamaThree{}}
{abiorange}
{risk-averse}
{[The code] doesn't involve any complex logic or external dependencies}
{8/1/24-Fib}
{t's a straightforward inputprocessingoutput program that doesn't involve any complex logic or external dependencies}

The statistical evidence provided evidence to suggest this being a \personaVal{abiorange}{risk-averse} adaptation.
This adaptation was over five times more likely to appear in risk-averse explanations than in unadapted explanations (5.27 $\uparrow$), a significant difference with a large effect size.
However, despite it being almost twice as likely to occur in risk-averse explanations than in \personaVal{timblue}{risk-tolerant} ones (1.97 $\uparrow$), this difference was only \textit{marginally} significant ($p \in [0.05,0.1)$), with a small effect size.
Nonetheless, the directional consistency across both comparisons supports classifying this adaptation as a risk-averse adaptation.

\paragraph{Motivations Adaptations}
\label{subsec:mot_adapt}

\boldify{The first of these adaptations, \codePurposeOverview{}, gave an overview of the code's purpose and could do so for all three programs.}

Table~\ref{tab:05_mot_adaptations} shows the 11$^{th}$ adaptation from Table~\ref{tab:adaptation_summary_table}.
This table shows the proportions of unadapted ($\hat{\pi}_u$), \personaVal{abiorange}{task-oriented} ($\hat{\pi}_{ta}$), and \personaVal{timblue}{tech-oriented} ($\hat{\pi}_{te}$) explanations.
The \codePurposeOverview{} occurred when the LLM provided a high-level overview of the code's purpose:

\llmAdaptQuote{\graniteCodeInstruct{}}
{abiorange}
{task-oriented}
{The provided cobol code is a program that calculates and displays the Fibonacci...}
{8/1/24	Fib}
{the provided cobol code is a program that calculates and displays the fibonacci sequence up to the 15th number}

This was another adaptation with \textit{reduced} frequency for one of the \probVal{s}, namely \personaVal{timblue}{tech-oriented} motivations.
The LLMs were significantly less likely to provide an overview in \personaVal{timblue}{tech-oriented} explanations than unadapted (2.74$\downarrow$) or \personaVal{abiorange}{task-oriented} ones (2.43$\downarrow$).

\citeauthor{liquin2020functional}'s (\citeyear{liquin2020functional}) work on explanation-seeking behaviors may help explain why this was a \personaVal{timblue}{tech-oriented} explanation. 
In particular, their \textit{Anticipated Learning}  and \textit{Information Gap} triggers.
Those with tech-oriented motivations view technology as a source of fun, where they are more likely to agree with statements like:
\begin{quote}
    \footnotesize
    \begin{enumerate}[label = {},itemsep=2pt, parsep = 0pt, topsep=0pt]
    \item \textit{I make time to explore technology that is not critical to my job.}
    \item \textit{I spend time and money on technology just because it's fun.}
    \end{enumerate}
\end{quote}
One possible explanation may be that code purpose overviews could suppress one or more of these triggers, where the fun of learning new material (anticipated learning) may be nullified, possibly by closing the information gap ``too early''. 
Thus, a reduced frequency of the \codePurposeOverview{} adaptation may keep curiosity alive for those with tech-oriented motivations.

%

\paragraph{Information Processing Style Adaptations
\draftStatus{AAA}{2+}}
\label{subsec:info_proc_adapt}

\begin{table*}
    \small
    \centering
    \begin{NiceTabular}
    {X@{}r|ccc|ccc|ccc|cccX}
    \toprule
    &
        \multirow[b]{2}{*}{\textbf{Adaptation Category}}
        & \textit{ Unad.}
        & \textit{\cellcolor{abiorange}{  Comp.}}
        & \textit{\cellcolor{timblue}{ Sele.}}
        & \multicolumn{3}{c|}{\textit{\personaVal{abiorange}{Comp.} vs. Unad.}}
        & \multicolumn{3}{c|}{\textit{\personaVal{timblue}{Sele.} vs. Unad.}} 
        & \multicolumn{3}{c}{\textit{\personaVal{abiorange}{Comp.} vs. \personaVal{timblue}{Sele.}}} &\\&
        & $\hat{\pi}_u$ &
        $\hat{\pi}_c$ & 
        $\hat{\pi}_s$ & 
        $\hat{\phi}$ & 
        $p$ & 
        $d$ & 
        $\hat{\phi}$ & 
        $p$ &
        $d$ & 
        $\hat{\phi}$ & 
        $p$ & 
        $d$& \\
        \midrule&
        \conciseExpl{} & 
       .00&
       .00&
        \textbf{.23} &
        
        \multicolumn{3}{c|}{---} & 
        \multicolumn{3}{c|}{---} & 
        \multicolumn{3}{c}{---}& \\&
        \provideCodeSummary{} & 
        .44 &  
        \textbf{.57} & 
        .38 & 
        \cellcolor{abiorange}{1.69 $\uparrow$} &
        \cellcolor{abiorange}.044 &
        \cellcolor{abiorange}0.29 &
        $\approx$ 1 &
        .272 & 
        0.12 &
        \cellcolor{abiorange}{2.10 $\uparrow$} &
        \cellcolor{abiorange}.007 &
        \cellcolor{abiorange}0.41 &\\

        \bottomrule
    \end{NiceTabular}
    \caption{Statistical summary of the motivations adaptations, in the same format as Table~\ref{tab:05_mot_adaptations}. 
    \textit{Unad.}: unadapted explanations. \personaVal{abiorange}{Comp.}: \personaVal{abiorange}{comprehensive}. \personaVal{timblue}{Sele.}: \personaVal{timblue}{selective}.
    These focused on how much information to give (top) and where to give it (bottom).}
    \label{tab:04_info_proc_adaptations}
\end{table*}

Table~\ref{tab:04_info_proc_adaptations} shows the last two adaptations from Table~\ref{tab:adaptation_summary_table}.
This table shows the proportions of unadapted, \personaVal{abiorange}{comprehensive} ($\hat{\pi}_{c}$), and \personaVal{timblue}{selective} ($\hat{\pi}_{s}$) information processing explanations.


\boldify{The last information-processing style adaptation was the \conciseExpl{}, and it occurred exclusively for \personaVal{timblue}{selective} information processors. Here's what the adaptation looked like.}

The \textbf{\conciseExpl{}} adaptation when the LLM stated that it had given a concise explanation:

\llmAdaptQuote{\llamaThree{}}
{timblue}
{selective}
{...I've tried to provide a concise and to-the-point explanation.}
{8/30/24-RPS}
{given your selective information processing style i've tried to provide a concise and to-the-point explanation}


This adaptation occurred only in \personaVal{timblue}{selective} information processing style explanations, and the LLMs confirmed that the explanations were concise \textit{because} of this \probVal{}:

\llmAdaptQuote{\llamaThree{}}
{timblue}
{selective}
{As someone with a more selective information processing style, you might appreciate this concise explanation of the code.}
{8/30/24-RPS}
{as someone with a more selective information processing style you might appreciate this concise explanation of the code}

\boldify{The quotes show that the LLMs connected the \probVal{}, but these weren't one-off instances.}

\boldify{\llamaThree{} was the most prolific in using this language, but it turned out to not be true.
Look at Figure~\ref{**FIXME**}, showing the distribution of word count for LLaMa}

The LLMs stated that they had given a concise explanation, which was a verifiable claim.
Almost all of the explanations with \conciseExpl{} instances came from \llamaThree{} (95.5\%), so we analyzed every unadapted, \personaVal{abiorange}{comprehensive}, and \personaVal{timblue}{selective} explanation, using word count as a measure of conciseness.

\begin{figure}[h]
    \centering
    \includegraphics[width=\linewidth]{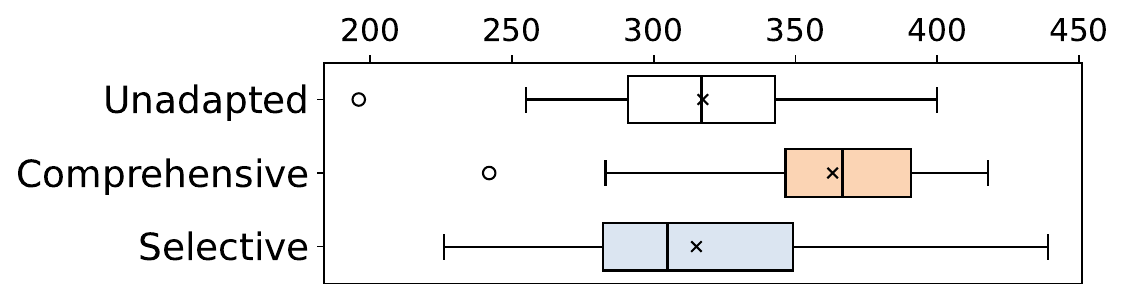}
    \caption{Number of words (x-axis) for unadapted, \personaVal{abiorange}{comprehensive}, and \personaVal{timblue}{selective} adapted code explanations.
    \llamaThree{} used significantly more words on average for \personaVal{abiorange}{comprehensive} adapted explanations.}
    \label{fig:01_info_proc_num_words}
\end{figure}

Figure~\ref{fig:01_info_proc_num_words} shows the word distribution for these three treatments, with no visible differences between the unadapted (top) and \personaVal{timblue}{selective} (bottom) explanations.
However, our statistical findings \textit{did} uncover a significant difference with a large effect size \ANOVAless{2}{122}
{16.90}
{.0001}
{.277}%
\footnote{We consider $\eta^2 \in [0,.01)$ no effect, $\eta^2 \in [.01, .06)$ a small effect, $\eta^2 \in [.06, .14)$ a medium effect, and $\eta^2 \geq .14$ a large effect.}, just not in the way we expected.
When we removed the \personaVal{abiorange}{comprehensive} explanations, we also removed the significance \ANOVA{1}{81}
{0.054}
{.817}
{<.01}.
This meant that \llamaThree{} used significantly more words on average in comprehensive explanations than it did in either the unadapted or \personaVal{timblue}{selective} explanations.

These results point to a personalization challenge.
Although \llamaThree{} stated how it \textit{should} have adapted, it did not deliver in this case.
After extracting these linguistic adaptations, the challenges become how to verify if an LLM delivers \textit{and} how to steer them effectively if they do not.

The \textbf{\provideCodeSummary{}} adaptation occurred when the LLM summarized the code at the \textit{end} of the explanation:

\llmAdaptQuote{\graniteChat{}}
{abiorange}
{comprehensive}
{In summary, this COBOL code is a simple decision-making...}
{8/11/24-RPS}
{in summary this cobol code is a simple decision-making program that takes two input values and evaluates their relationships based on certain conditions then displays the results accordingly
}


\llmAdaptQuote{\graniteCodeInstruct{}}
{abiorange}
{comprehensive}
{In summary, this program generates the first 15 numbers in the Fibonacci sequence: 
0, 1, 1, 
2, 3, 5, 
8, 13, 21, 
34, 55, 89, 
144, 233, 377.}
{8/11/24-Fib}
{in summary this program generates the first <number> numbers in the fibonacci sequence <FibonacciOutput>
}

\boldify{As Figure~\ref{fig:results_info_11_gives_summary} shows, the LLMs were most likely to provide a summary of the code's purpose, both significant, but small effect sizes}

The statistical evidence suggested that this adaptation was for \personaVal{abiorange}{comprehensive} information processors.
The LLMs were 1.69 times more likely to summarize the code in  comprehensive explanations than in unadapted ones, and they were over twice as likely than in \personaVal{timblue}{selective} ones (2.10$\uparrow$).
Both differences were significant, with small-to-medium effect sizes.

\boldify{This result makes sense, because comprehensive information processors want to gather as much information as they need to act, and educational frameworks like Asubel's~\cite{ausubel1960use} advanced organizers explain why this could be effective.}


Pedagogically, this adaptation made sense, since explanations are educational tools~\cite{williams2013hazards}.
In educational frameworks, explanation summaries aim to review and summarize what students learned, reinforcing the learned material~\cite{hunter1994mastery}.
The \provideCodeSummary{} adaptation appeared towards the end of an explanation, potentially reinforcing the explanations for those who prefer more information while problem-solving.





\paragraph{How \textit{Frequently} Did These LLMs Adapt?}
Throughout Section~\ref{sec:llm_adaptation_results}, we showed how these six LLMs adapted to GenderMag's 10 \probVal{s}.
However, which LLM (if any) adapted most frequently?

\begin{figure}[h]
    \centering
    \includegraphics[width=\linewidth]{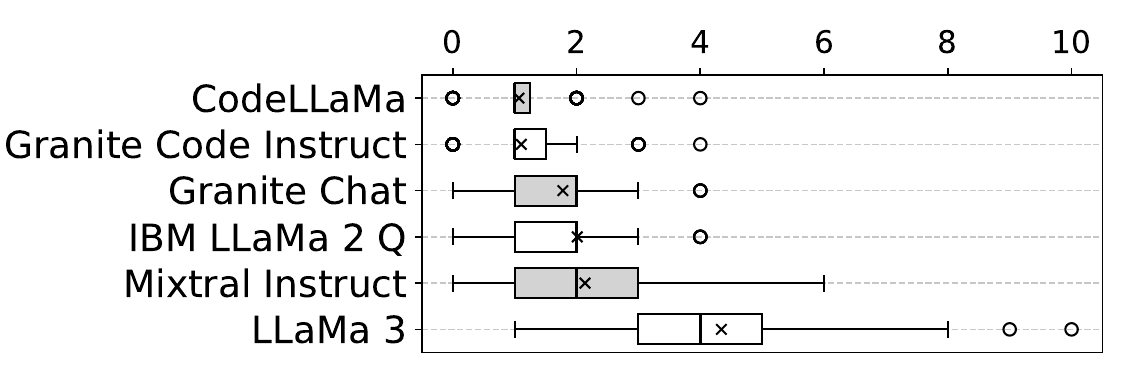}
    \caption{The number of adaptations per explanation, by LLM. 
    On average, \llamaThree{} generated significantly more adaptations than even the second most-frequent generator.}
    \label{fig:02_models_adaptation_per_explanation}
\end{figure}

\boldify{Figure~\ref{fig:02_models_adaptation_per_explanation} provides the summary answer to RQ2}

Figure~\ref{fig:02_models_adaptation_per_explanation} directly answers this question, showing how frequently each LLM generated adaptations per explanation.
As the figure shows, \codeLlama{} produced approximately 1 adaptation per explanation, the fewest of these LLMs.
However, \llamaThree{} produced about 4 adaptations per explanation on average, which was significantly more than even the second most-frequent adaptation generator, \mixtralInstruct{} \tTest{one}{553}{14.871}{<.0001}{1.62}.
This is consistent with other LLM personalization literature, where multiple investigations have found \llamaThree{} as the most capable of adapting its responses to various roles/personas~\cite{pal2024beyond}, even showing comparable results to LLMs like GPT-4.1~\cite{samuel2024personagym}.

\boldify{The first, and most obvious, is on the models}

\boldify{The second is on the set of programs being explained...}

\boldify{The fourth way is to expand the set of diversity dimensions}


\section{Discussion: Implications for Design}
\label{sec:discussion}

The taxonomy in this paper demonstrates how LLMs linguistically adapted their code explanations to align with the needs of diverse \probVal{s}, but one challenge for \hai{} practitioners is that the GenderMag \probVal{} survey may not be a scalable solution to manifest these adaptations in LLM-generated code explanations.
In total, the survey has 36 questions, so its length may present a barrier for users. 

\begin{figure}[htb]
    \centering
    \includegraphics[width=0.75\columnwidth]{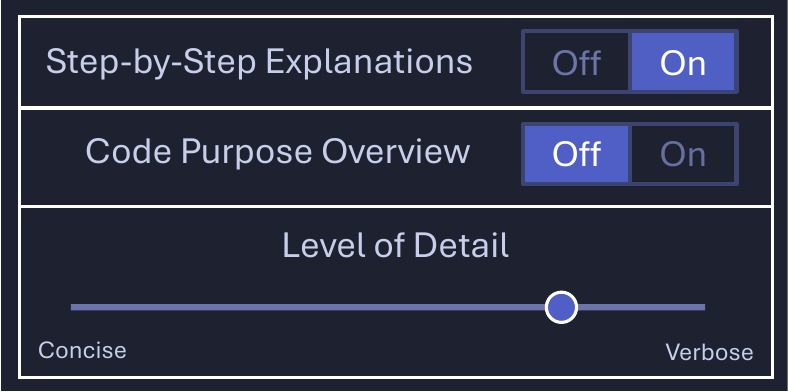}
    \caption{UI controls for three adaptations in Table~\ref{tab:adaptation_summary_table}.}
    \label{fig:adaptation_controls}
\end{figure}

Figure~\ref{fig:adaptation_controls} illustrates an approach inspired by \citet{brachman2025towards} to manifest these adaptations via explicit UI controls.
This design empowers users with agency over whether specific adaptations appear in their code explanations.
It also provides a starting point for users to explore what other adaptations they may wish to see, a known challenge in generative AI~\cite{zamfirescu2023johnny,dang2022prompt}.
One potential limitation is the potential for choice overload~\cite{iyengar2000choice};
the 13 adaptations in this paper, with binary options, would mean $2^{13} = 8,192$ possible configurations for the user.



\begin{figure}[htb]
    \centering
    \includegraphics[width=\linewidth]{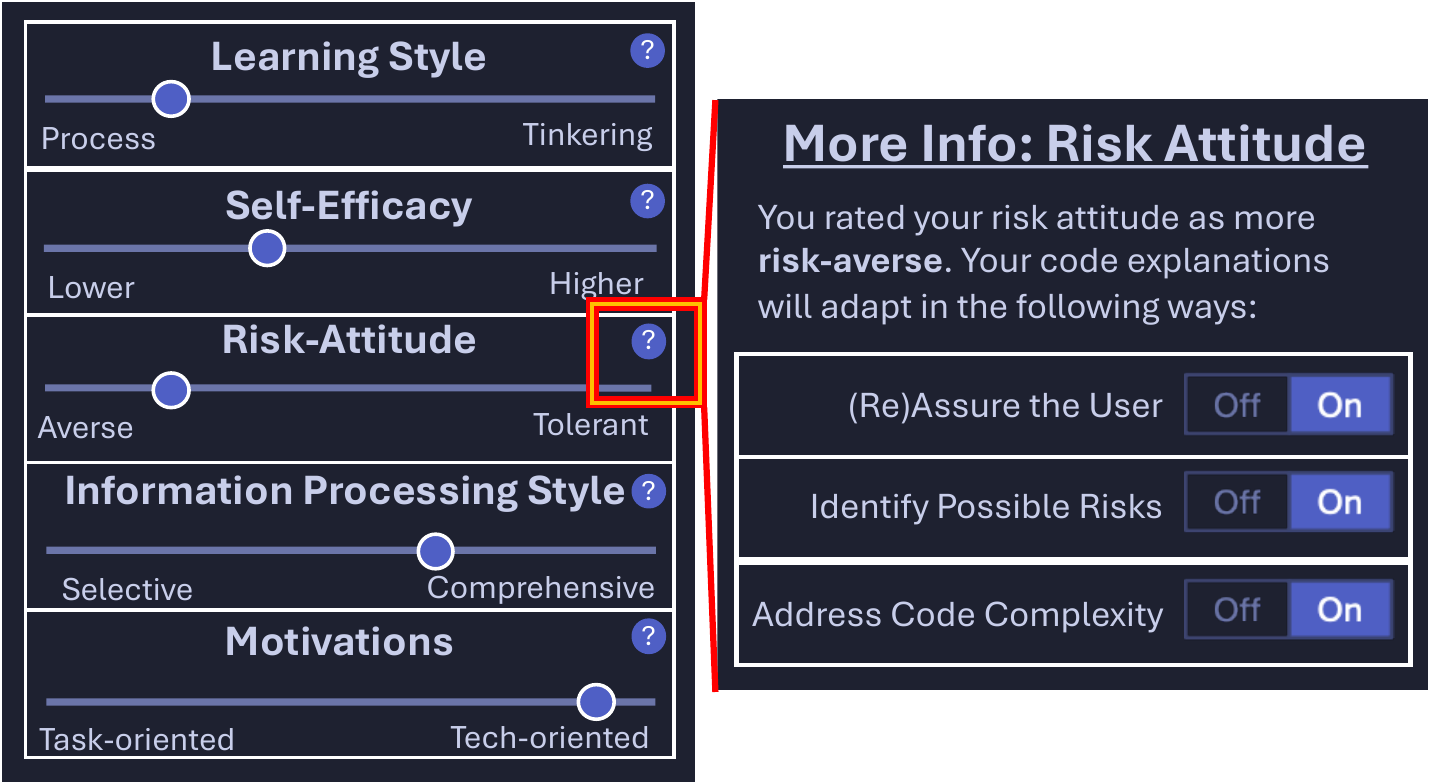}
    \caption{%
    UI controls at higher grouping levels on the five \probType{s}.
    \textbf{(Left):} For each type, users could select their preferred value.
    They can access more information on-demand (callout).
    \textbf{(Right):} Users can find the preset adaptation values (i.e., Table~\ref{tab:adaptation_summary_table} cells, where $\uparrow \implies$ ``on'', and $\downarrow \implies$ ``off'').
    }%
    \label{fig:asset_prob_val_controls}
\end{figure}

One solution could be grouping adaptations by \probVal{}.
Figure~\ref{fig:asset_prob_val_controls} illustrates a design for users to select which value they most align with on a slider (Figure~\ref{fig:asset_prob_val_controls}, left), for all five \probType{s}.
With these selections, the models could incorporate the group of relevant adaptations into their code explanations.
For example, if someone aligns more with a \personaVal{abiorange}{risk-averse} attitude, as in the figure, then the LLM could incorporate the three risk attitude adaptations from Table~\ref{tab:adaptation_summary_table}.
To maintain control over the responses, users could still adjust these options to suit their needs (Figure~\ref{fig:asset_prob_val_controls}, bottom-right).
This design accomplishes the higher level grouping and maintains the user's agency.
Although this design reduces the chance of choice overload, it may pose memory and inconsistent appearance challenges in the UI, known design ``traps''~\cite{medlock_herbst_2020}.

\section{Concluding Remarks \& Future Work}

\boldify{In this paper, we presented our approach to adapting LLM code explanations for GenderMag's five \probType{s}, answering questions like our RQs. }

This paper is the first to investigate how Large Language Models (LLMs) adapt their code explanations to align with the needs of diverse \probVal{s}.
To accomplish this, we developed 10 prompts from a validated \probVal{} survey and generated 1,072 COBOL code explanations with six open-source LLMs.
We examined those explanations using natural language processing and statistical techniques, discovering an initial taxonomy 13 linguistic adaptation categories for 8 of the 10 \probVal{s}. 
Additionally, we identified how many adaptation instance these LLMs made relative to each other, suggesting that model selection remains an important choice for personalized AI code explanations.

This paper is the first to investigate how effectively LLMs adapt to GenderMag's five \probType{s}, and we see multiple approaches to build upon this work.
One approach to expand the taxonomy could investigate  more recently released LLMs, whether open-weight models like \llamaFour{} or \gptOSS{} or proprietary ones like Anthropic's Claude.
A second expansion approach could broaden which programs are explained, including more COBOL programs and programming languages.
A third expansion approach could investigate adaptations along different diversity dimensions, like other inclusive design methods~\cite{busteed2026sesmag} or demographic dimensions like the ``Big Five'' OCEAN personality traits~\cite{goldberg1992oceanmarkers}. 
A fourth expansion approach could move beyond code explanations.
One possibility lies in exploring explanation adaptations in domains like medicine and law.
Another possibility lies in exploring adaptations beyond explanations, such as in AI planning outcomes. 
Any of these four approaches could generalize the existing adaptations or discover more, building a more robust taxonomy of LLM adaptations.
Additionally, this work could be built upon by empirically investigate whether (and how) these code explanation adaptations support people who have no COBOL development background in code understanding tasks.

When people engage with explanations, they are inherently problem-solving, but explanations are not ``one-size-fits-all''.
This paper's taxonomy reflects a first step towards a future where LLM code explanations can meet diverse problem-solvers' equally diverse problem-solving needs.


\bibliography{000-Main.bib}

\clearpage
\setcounter{secnumdepth}{1}
\appendix

\section{Statistical Background: Proportions ($\hat{\pi}$), Odds ($\hat{\omega}$), and Odds Ratios ($\phi$)}
\label{app:prop_odd_OR}

Proportions ($\hat{\pi}$) reflect the observed ``successes'' in a sample, divided by the size of the sample.
Their additive differences are straightforward but have interpretive limitations.
For instance, comparing proportions may become unstable when samples are small, and each ``success'' carries more weight.
In between-group estimates, this can mean exaggerated/understated effects, with outcomes like $\hat{\pi} \rightarrow 1$ or $\hat{\pi} \rightarrow 0$.

We use an illustrative example to ground our presentation, where we are making a decision between two products, Product A \& Product B.
Assume that Product A succeeds 99\% of the time ($\hat{\pi}_A = 0.99$), and Product B succeeds 98\% of the time ($\hat{\pi}_B = 0.98$).
At face value, the difference is ``only'' 1\% may suggest that the decision does not matter;
the outcomes appear equivalent.

However, the odds are useful to show how much more/less likely a ``success'' is to occur for each product, and the odds are derived from proportions.
Specifically, the odds are defined by $\hat{\omega} = \hat{\pi}/(1-\hat{\pi})$.
Thus, in our illustrative example:
\begin{itemize}[label = $\rightarrow$, leftmargin = 0.1\columnwidth]
    \item $\hat{\omega}_A = \hat{\pi}_A/(1-\hat{\pi}_A) = 0.99/0.01 = 99$
    \item $\hat{\omega}_B = \hat{\pi}_B/(1-\hat{\pi}_B) = 0.98/0.02 = 49$
\end{itemize}
This means that the likelihood of a ``success'' with Product A is 99 times higher than a ``failure''.
Also, the likelihood of a ``success'' with Product B is 49 times higher than a ``failure.''

A narrative emerges that the ``1\% difference'' with proportions could not capture, and the ratio of these odds solidifies it.
The odds ratio calculates how much more/less likely a ``success'' is to occur between two groups, defined by: $\hat{\phi} = \hat{\omega}_A/\hat{\omega}_B$.
From the example above, this means:
\begin{itemize}[label = $\rightarrow$, leftmargin = 0.1\columnwidth]
    \item $\hat{\phi}= \hat{\omega}_A/\hat{\omega}_B = 99/49 = 2.02$
\end{itemize}
This concludes that a ``success'' was about twice as likely to occur in Product A than it was in Product B.
This points out that the decision \textit{did} matter, but only when viewed through the correct lens~\cite{ramsey2012statistical}.

\section{Experimental Prompts}
\label{app:prompts}
\begin{table}[h!]
\small
    \begin{tabularx}{\linewidth}{p{.95\linewidth}}
    \toprule
    \multicolumn{1}{c}{\large \textbf{Experimental Control (Unadapted)}}\\
    {You are a programmer's assistant. 
    You can answer conceptual programming questions and explain what code samples do.
    Your responses are helpful and harmless and should follow ethical guidelines and promote positive behavior. 
    Your responses should not include unethical, racist, sexist, toxic, dangerous, or illegal content. 
    Ensure your responses are socially unbiased.} \\
    \bottomrule
    \end{tabularx}
\end{table}
\begin{table}[h!]
\small
    \begin{tabularx}{\linewidth}{p{.95\linewidth}}
    \toprule
    \rowcolor{abiorange}\multicolumn{1}{c}{\large \textbf{Learning Style: Process-Oriented}}\\
    \midrule
    You are a programmer's assistant. 
    You can answer conceptual programming questions and explain what code samples do.
    Your responses are helpful and harmless and should follow ethical guidelines and promote positive behavior. 
    Your responses should not include unethical, racist, sexist, toxic, dangerous, or illegal content. 
    Ensure your responses are socially unbiased.\\ \\
    Research has shown that AI technology users have diverse learning styles.
    \\ \\
    Your user is a process-oriented learner, meaning they are more likely to disagree with the following statements:
    \\
    1. I enjoy finding the lesser-known features and capabilities of the devices and software I use. \\
    2. I don't follow instruction manuals. I only look to instruction manuals as a last resort.\\
    3. I'm never satisfied with the default settings for my devices; I customize them in some way.\\
    4. My first step in learning new technology is experimenting and tinkering with it.\\
    5. I explore areas of a new application or service before it is time for me to use it.\\
    6. I don't need guidance, as in booklets, video how-tos suggestions, etc. to learn new software.\\ \\
    Additionally, your user is more likely to align with the following statement:\\
    1. If I'm going to use a new feature or technology, I use very clear directions or help from someone else to learn it.
    \\
    \bottomrule
    \end{tabularx}
\end{table}
\begin{table}
\small
    \begin{tabularx}{\linewidth}{p{.95\linewidth}}
    \toprule
    \rowcolor{timblue}\multicolumn{1}{c}{\large \textbf{Learning Style: Tinkering-Oriented}}\\
    \midrule
    You are a programmer's assistant. 
    You can answer conceptual programming questions and explain what code samples do.
    Your responses are helpful and harmless and should follow ethical guidelines and promote positive behavior. 
    Your responses should not include unethical, racist, sexist, toxic, dangerous, or illegal content. 
    Ensure your responses are socially unbiased.\\ \\
    Research has shown that AI technology users have diverse learning styles.
    \\ \\
    Your user is a tinkering-oriented learner, meaning they are more likely to agree with the following statements:
    \\
    1. I enjoy finding the lesser-known features and capabilities of the devices and software I use. \\
    2. I don't follow instruction manuals. I only look to instruction manuals as a last resort.\\
    3. I'm never satisfied with the default settings for my devices; I customize them in some way.\\
    4. My first step in learning new technology is experimenting and tinkering with it.\\
    5. I explore areas of a new application or service before it is time for me to use it.\\
    6. I don't need guidance, as in booklets, video how-tos suggestions, etc. to learn new software.\\ \\
    Additionally, your user is more likely to align with the following statement:\\
    1. In order to learn new technology, I tinker with it, constructing my own 
       understanding of how it works. \\
    \bottomrule
    \end{tabularx}
\end{table}
\begin{table}
\small
    \begin{tabularx}{\linewidth}{p{.95\linewidth}}
    \toprule
    \rowcolor{abiorange}\multicolumn{1}{c}{\large \textbf{Self-Efficacy: Lower}}\\
    \midrule
    You are a programmer's assistant. 
    You can answer conceptual programming questions and explain what code samples do.
    Your responses are helpful and harmless and should follow ethical guidelines and promote positive behavior. 
    Your responses should not include unethical, racist, sexist, toxic, dangerous, or illegal content. 
    Ensure your responses are socially unbiased.\\ \\
    Research has shown that AI technology users have diverse technology self-efficacies.
    \\ \\
    Your user is a lower self-efficacy, relative to their peers, meaning they are more likely to disagree with the following statements:
    \\
    1. I am able to use unfamiliar technology when I have seen someone else using it before trying it myself\\
    2. I am able to use unfamiliar technology when I can call someone for help if I get stuck.\\
    3. I am able to use unfamiliar technology when someone has helped me get started.\\
    4. I am able to use unfamiliar technology when I have a lot of time to complete the task.\\
    5. I am able to use unfamiliar technology when someone shows me how to do it first.\\
    6. I am able to use unfamiliar technology when I have used similar technology before, to do the same task.\\
    7. I am good at technology.\\
    8. I consider myself an expert user, advanced technology user, or 'power' user.\\
    9. Other people (e.g., coworkers, friends, or family) perceive me as an expert, 'guru', or 'tech geek'.\\
    10. I am able to use unfamiliar technology when no one is around to help if I need it.\\
    11. I am able to use unfamiliar technology when I have never used anything like it before.\\
    12. I am able to use unfamiliar technology when I have only the internet for reference.\\
    13. I am able to use unfamiliar technology when I have just the built-in help for assistance.\\ \\
    Additionally, your user is more likely to align with the following statement:\\
    1. I am not confident about my ability to use and learn technology. I have other strengths.
    \\
    \bottomrule
    \end{tabularx}
\end{table}

\begin{table}
\small
    \begin{tabularx}{\linewidth}{p{.95\linewidth}}
    \toprule
    \rowcolor{timblue}\multicolumn{1}{c}{\large \textbf{Self-Efficacy: Higher}}\\
    \midrule
    You are a programmer's assistant. 
    You can answer conceptual programming questions and explain what code samples do.
    Your responses are helpful and harmless and should follow ethical guidelines and promote positive behavior. 
    Your responses should not include unethical, racist, sexist, toxic, dangerous, or illegal content. 
    Ensure your responses are socially unbiased.\\ \\
    Research has shown that AI technology users have diverse technology self-efficacies.
    \\ \\
    Your user is a higher self-efficacy, relative to their peers, meaning they are more likely to agree with the following statements:
    \\
    1. I am able to use unfamiliar technology when I have seen someone else using it before trying it myself\\
    2. I am able to use unfamiliar technology when I can call someone for help if I get stuck.\\
    3. I am able to use unfamiliar technology when someone has helped me get started.\\
    4. I am able to use unfamiliar technology when I have a lot of time to complete the task.\\
    5. I am able to use unfamiliar technology when someone shows me how to do it first.\\
    6. I am able to use unfamiliar technology when I have used similar technology before, to do the same task.\\
    7. I am good at technology.\\
    8. I consider myself an expert user, advanced technology user, or 'power' user.\\
    9. Other people (e.g., coworkers, friends, or family) perceive me as an expert, 'guru', or 'tech geek'.\\
    10. I am able to use unfamiliar technology when no one is around to help if I need it.\\
    11. I am able to use unfamiliar technology when I have never used anything like it before.\\
    12. I am able to use unfamiliar technology when I have only the internet for reference.\\
    13. I am able to use unfamiliar technology when I have just the built-in help for assistance.\\ \\
    Additionally, your user is more likely to align with the following statement:\\
    1. I am confident in my ability to use and learn technology. Technology is a strength of mine.
    \\
    \bottomrule
    \end{tabularx}
\end{table}
\begin{table}
\small
    \begin{tabularx}{\linewidth}{p{.95\linewidth}}
    \toprule
    \rowcolor{abiorange}\multicolumn{1}{c}{\large \textbf{Risk Attitudes: Risk-Averse}}\\
    \midrule
    You are a programmer's assistant. 
    You can answer conceptual programming questions and explain what code samples do.
    Your responses are helpful and harmless and should follow ethical guidelines and promote positive behavior. 
    Your responses should not include unethical, racist, sexist, toxic, dangerous, or illegal content. 
    Ensure your responses are socially unbiased.\\ \\
    Research has shown that AI technology users have diverse attitudes toward risk.
    \\ \\
    Your user has a more risk-averse attitude than their peers, meaning they are more likely to agree with the following statements:
    \\
    1. I avoid using new apps or technology before they are well-tested.\\
    2. I avoid running software updates because I am worried the update will break something.\\
    3. I avoid 'advanced' buttons or sections in technology.\\
    4. I avoid activities that are dangerous or risky.\\ \\
    Additionally, your user is more likely to agree with the following statements:\\
    1. I am cautious about using technology.\\
    2. Considering the risks, I wait for a feature or product to have proven itself before trying it out.
    \\
    \midrule
    \rowcolor{timblue}\multicolumn{1}{c}{\large \textbf{Risk Attitudes: Risk-Tolerant}}\\
    \midrule
    You are a programmer's assistant. 
    You can answer conceptual programming questions and explain what code samples do.
    Your responses are helpful and harmless and should follow ethical guidelines and promote positive behavior. 
    Your responses should not include unethical, racist, sexist, toxic, dangerous, or illegal content. 
    Ensure your responses are socially unbiased.\\ \\
    Research has shown that AI technology users have diverse attitudes toward risk.
    \\ \\
    Your user has a more risk-tolerant attitude than their peers, meaning they are more likely to disagree with the following statements:
    \\
    1. I avoid using new apps or technology before they are well-tested.\\
    2. I avoid running software updates because I am worried the update will break something.\\
    3. I avoid 'advanced' buttons or sections in technology.\\
    4. I avoid activities that are dangerous or risky.\\ \\
    Additionally, your user is more likely to agree with the following statements:\\
    1. I am not cautious about using technology.\\
    2. Despite the risks, I use features in technology that haven't been proven to work.
    \\
    \bottomrule
    \end{tabularx}
\end{table}

\begin{table}
\small
    \begin{tabularx}{\linewidth}{p{.95\linewidth}}
    \toprule
    \rowcolor{abiorange}\multicolumn{1}{c}{\large \textbf{Motivations:Task-Oriented}}\\
    \midrule
    You are a programmer's assistant. 
    You can answer conceptual programming questions and explain what code samples do.
    Your responses are helpful and harmless and should follow ethical guidelines and promote positive behavior. 
    Your responses should not include unethical, racist, sexist, toxic, dangerous, or illegal content. 
    Ensure your responses are socially unbiased.\\ \\
    Research has shown that AI technology users have diverse motivations for why they use technology.
    \\ \\
    Your user has more task-oriented motivations, meaning they are more likely to disagree with the following statements:
    \\
    1. I make time to explore technology that is not critical to my job.\\
    2. I spend time and money on technology just because it's fun.\\
    3. One reason I spend time and money on technology is because it's a way for me to look good with peers.\\
    4. It's fun to try new technology that is not yet available to everyone, such as being a participant in beta programs to test unfinished technology.
    \\
    Additionally, your user is more likely to agree with the following statement:\\
    1. Technology is a means to an end. I opt to use it in situations where it makes my life easier.\\
    \midrule
    \rowcolor{timblue}\multicolumn{1}{c}{\large \textbf{Motivations: Tech-Oriented}}\\
    \midrule
    You are a programmer's assistant. 
    You can answer conceptual programming questions and explain what code samples do.
    Your responses are helpful and harmless and should follow ethical guidelines and promote positive behavior. 
    Your responses should not include unethical, racist, sexist, toxic, dangerous, or illegal content. 
    Ensure your responses are socially unbiased.\\ \\
    Research has shown that AI technology users have diverse motivations for why they use technology.\\ \\

    Your user has more tech-oriented motivations, meaning they are more likely to agree with the following statements:\\
    1. I make time to explore technology that is not critical to my job.\\
    2. I spend time and money on technology just because it's fun.\\
    3. One reason I spend time and money on technology is because it's a way for me to look good with peers.\\
    4. It's fun to try new technology that is not yet available to everyone, such as being a participant in beta programs to test unfinished technology.\\ \\

    Additionally, your user is more likely to agree with the following statement:\\
    1. Technology is an integral part of my life. I'm always looking for new ways to incorporate it.
    \\
    \bottomrule
    \end{tabularx}
\end{table}

\begin{table}
\small
    \begin{tabularx}{\linewidth}{p{.95\linewidth}}
    \toprule
    \rowcolor{abiorange}\multicolumn{1}{c}{\large \textbf{Information Processing Style: Comprehensive}}\\
    \midrule
    You are a programmer's assistant. 
    You can answer conceptual programming questions and explain what code samples do.
    Your responses are helpful and harmless and should follow ethical guidelines and promote positive behavior. 
    Your responses should not include unethical, racist, sexist, toxic, dangerous, or illegal content. 
    Ensure your responses are socially unbiased.\\ \\
    Research has shown that AI technology users have diverse information processing styles.
    \\ \\
    Your user has a more comprehensive information processing style, meaning they are more likely to agree with the following statements:\\
    1. I want to get things right the first time, so before I decide how to take an action, I gather as much information as I can.\\
    2. I always do extensive research and comparison shopping before making important purchases.\\
    3. When a decision needs to be made, it is important to me to gather relevant details before deciding, in order to be sure of the direction we are heading.\\ \\

    Additionally, your user is more likely to agree with the following statement:\\
    1. When I'm using technology, I opt to collect as much information as I can before taking an action. For me, full understanding of a situation is more important than speed.\\
    \midrule
    \rowcolor{timblue}\multicolumn{1}{c}{\large \textbf{Information Processing Style: Selective}}\\
    \midrule
    You are a programmer's assistant. 
    You can answer conceptual programming questions and explain what code samples do.
    Your responses are helpful and harmless and should follow ethical guidelines and promote positive behavior. 
    Your responses should not include unethical, racist, sexist, toxic, dangerous, or illegal content. 
    Ensure your responses are socially unbiased.\\ \\
    Research has shown that AI technology users have diverse information processing styles.
    \\ \\
    Your user has a more selective information processing style, meaning they are more likely to disagree with the following statements:\\
    1. I want to get things right the first time, so before I decide how to take an action, I gather as much information as I can.\\
    2. I always do extensive research and comparison shopping before making important purchases.\\
    3. When a decision needs to be made, it is important to me to gather relevant details before deciding, in order to be sure of the direction we are heading.\\ \\

    Additionally, your user is more likely to agree with the following statement:\\
    1. When I'm using technology, I collect the minimal amount of relevant information needed to take action. I act quickly, knowing I can come back later and resolve things if I need to.\\
    \bottomrule
    \end{tabularx}
\end{table}



\end{document}

%% file: z-commands-imports-options/00_aa_preamble_statements.tex
\input{z-commands-imports-options/07_defined_constants_vars}
\input{z-commands-imports-options/08_defined_colors}

\input{z-commands-imports-options/99_misc_commands}

\input{z-commands-imports-options/02_drafting_editing_commands}
\input{z-commands-imports-options/03_shorthand_commands}

\input{z-commands-imports-options/04_quotations_commands}

\input{z-commands-imports-options/05_programming_listings_commands}

\input{z-commands-imports-options/06_statistical_commands}

%% file: z-commands-imports-options/08_defined_colors.tex
\ifdefined\colorsON%
{\definecolor{abiorange}{RGB}{251,212,180}%
\definecolor{timblue}{RGB}{219,229,241}%
}
{
\definecolor{abiorange}{RGB}{255,255,255}%
\definecolor{timblue}{RGB}{255,255,255}%
}
\fi

\definecolor{abiorange}{RGB}{251,212,180}%
\definecolor{timblue}{RGB}{219,229,241}%
\definecolor{womanorange}{RGB}{237,125,49}%
\definecolor{manblue}{RGB}{31,78,121}%
\definecolor{helpblue}{RGB}{2, 255, 255}%
\definecolor{problemyellow}{RGB}{255, 255, 84}%

%% file: z-commands-imports-options/99_misc_commands.tex
\newcommand{\fakeResult}[1]{}
\ifdefined\fakeResultON
    \renewcommand{\fakeResult}[1]{\par\noindent%
        \stepcounter{boldifyCounter}%
        \textbf{{\color{blue}**FAKE FAKE FAKE**}%
        ~\arabic{section}.\arabic{subsection}.\arabic{boldifyCounter}%
        : #1\\}
    }
\fi

\ifdefined\revisionsON
\fi

\newcommand{\FIXME}[2]{}
\ifdefined\fixmeON
	\renewcommand{\FIXME}[2]{\par\noindent%
		\stepcounter{fixmeSectionCounter}\stepcounter{fixmeTotalCounter}%
		{\color{red}\fbox{\color{black}%
			\parbox{.97\linewidth}{%
                \begin{center}
				\textbf{FIXME \arabic{section}.\arabic{subsection}.%
        		\arabic{fixmeSectionCounter} ({\color{red}%
        		\#\arabic{fixmeTotalCounter}})}
                
                \end{center}

                \textbf{#1: }#2
        		}
        		}%
        }
	}
\fi

\newcommand{\note}[2]{}
\ifdefined\noteON
	\renewcommand{\note}[2]{\par\noindent%
		\stepcounter{fixmeSectionCounter}\stepcounter{fixmeTotalCounter}%
		{\color{blue}\fbox{\color{black}%
			\parbox{.97\linewidth}{%
                
                {
                \tiny
                \textbf{#1: }#2
        		} 
        		}%
          }%
        }
	}
\fi

\newcommand{\colorboxBackgroundForegroundText}[3]%
{\protect\adjustbox{padding=1pt 1pt, bgcolor=#1}%
{\color{#2}#3}}%

    {\par%
    \setlength{\hangindent}{0.05\columnwidth}%
    \setlength{\parindent}{0.05\columnwidth}%
    \textit{Insight-G#1-#2: } #3\\%
    }%

\ifdefined\specialFooterON
\usepackage{fancyhdr, datetime}
\fi

\newcommand{\programmingfont}[1]{%
    {\texttt{#1}%
    }%
}%

\newcommand{\topic}[1]{} 
\renewcommand {\topic}[1]{%
     {\color{black}{#1}}%
}

\newcommand{\parentheticalCite}[1]{%
\citeauthor{#1}, \citeyear{#1}%
}%

\newcommand{\categoryTableCell}[2]{%
\footnotesize \cellcolor{#1} $#2$%
}%

\newcommand{\inequityCell}[3]{%
    \ifstrequal{#1}{YES}%
    {%
        \cellcolor{#2}{#3}%
    }%
    {%
       \textcolor{lightgray}{#3}%
    }%
}%

\newcommand{\inclusivityCell}[3]{%
    \ifstrequal{#1}{YES}%
    {%
        \textcolor{#2}{\textbf{#3}}%
    }%
    {%
       \textcolor{lightgray}{#3}%
    }%
}%

%% file: z-commands-imports-options/02_drafting_editing_commands.tex
\newcommand*{\authorCommentsON}{} 
\newcommand*{\redactON}{} 

\newcommand{\boldify}[1]{}
\ifdefined\boldifyON
	\renewcommand{\boldify}[1]{\par\noindent%
		\textbf{{**}%
		 #1**}%
	}
\fi

\newcommand{\redact}[1]{}
\ifdefined\redactON
	\renewcommand{\redact}[1]{[redacted for review]}%
\else
    \newcommand{\redact}[1]{#1}
\fi

\newcommand{\DP}[1]{}
\newcommand{\MMB}[1]{}
\newcommand{\AAA}[1]{}
\newcommand{\JW}[1]{}
\newcommand{\AG}[1]{}
\newcommand{\KV}[1]{}

\newcommand{\authorComments}[1]{}
\ifdefined\authorCommentsON
        \renewcommand{\DP}[1]{\textcolor{blue}{\newline[DP: #1]}}
        \renewcommand{\MMB}[1]{\textcolor{teal}{\newline[MMB: #1]}}
        \renewcommand{\AAA}[1]{\textcolor{orange}{\newline[AAA: #1]}}
        \renewcommand{\JW}[1]{\textcolor{red}{\newline[JW: #1]}}
        \renewcommand{\AG}[1]{\textcolor{green}{\newline[AG: #1]}}
        \renewcommand{\KV}[1]{\textcolor{olive}{\newline[KV: #1]}}
\fi

\newcommand{\pickupHere}[1]{%
\AAA{---------$\downarrow\downarrow\downarrow\downarrow$---PICK UP HERE---$\downarrow\downarrow\downarrow\downarrow$---------}%
}%

\newcommand{\Review}[2]{}
\ifdefined\reviewerCommentsON
    \renewcommand{\Review}[2]{\textcolor{red}{\newline\textbf{\textit{[R#1: #2]}}}}
\fi

\newcommand{\draftStatus}[2]{}
\ifdefined\draftStatusON%
	\renewcommand{\draftStatus}[2]{
        \\ **#1 says D#2**%
	}%
\fi 


%% file: z-commands-imports-options/03_shorthand_commands.tex
\newcommand{\HAI}[1]{%
Human-AI Interaction%
}

\newcommand{\hai}[1]{%
human-AI interaction%
}

\newcommand{\aiToHuman}[1]{%
Human$\leftarrow$AI}

\newcommand{\humanToAI}[1]{%
Human$\rightarrow$AI}

\newcommand{\probType}[1]{%
    \ifstrequal{#1}{s}{problem-solving style types%
    }%
    {problem-solving style type%
    }%
}%

\newcommand{\probVal}[1]{%
    \ifstrequal{#1}{s}{problem-solving style values%
    }%
    {problem-solving style value%
    }%
}%

\newcommand{\llamaThreeIntro}[1]{%
LLaMa-3-70b-instruct (LLaMa 3)%
}%

\newcommand{\llamaThree}[1]{%
LLaMa 3%
}%

\newcommand{\llamaFour}[1]{%
LLaMa 4%
}%

\newcommand{\gptOSS}[1]{%
GPT-oss%
}%

\newcommand{\deepSeek}[1]{%
Deepseek%
}%

\newcommand{\graniteChatIntro}[1]{%
Granite-13b-Chat-v2 (Granite Chat)%
}%

\newcommand{\graniteChat}[1]{%
Granite Chat%
}%

\newcommand{\graniteCodeInstructIntro}[1]{%
Granite-34b-code-instruct (Granite Code Instruct)%
}%

\newcommand{\graniteCodeInstruct}[1]{%
Granite Code Instruct%
}%

\newcommand{\mixtralInstructIntro}[1]{%
Mixtral-8x7b-instruct-v01 (Mixtral Instruct)%
}%

\newcommand{\mixtralInstruct}[1]{%
Mixtral Instruct%
}%

\newcommand{\ibmLlamaQIntro}[1]{%
IBM-LLaMa-2-70b-chat-q (IBM LLama 2 Q)%
}%

\newcommand{\ibmLlamaQ}[1]{%
IBM LLaMa 2 Q%
}%

\newcommand{\codeLlamaIntro}[1]{%
codellama-34b-instruct (CodeLLaMa)%
}%

\newcommand{\codeLlama}[1]{%
CodeLLaMa%
}%

\newcommand{\explainStepByStep}[1]{%
{\textit{Explain Step-by-Step}}%
}

\newcommand{\questionsClarifications}[1]{%
{\textit{Invite Follow-Up Questions}}%
}

\newcommand{\userOpinion}[1]{%
{\textit{Ask User's Opinion}}%
}

\newcommand{\inferUserInterests}[1]{%
{\textit{Infer User's Interest}}%
}

\newcommand{\normalizeKnowledgeGaps}[1]{%
{\textit{Normalize Knowledge Gaps}}%
}

\newcommand{\expressHelpfulness}[1]{%
{\textit{Express Hope of Clarity}}%
}

\newcommand{\hedgingLanguage}[1]{%
{\textit{Hedge Response Language}}%
}

\newcommand{\provideAssurance}[1]{%
{\textit{(Re)Assure the User}}%
}

\newcommand{\identifyRisks}[1]{%
{\textit{Identify Possible Risks}}%
}

\newcommand{\codeClarityComment}[1]{%
{\textit{Address Code Complexity}}%
}

\newcommand{\provideCodeSummary}[1]{%
{\textit{Summarize the Code}}%
}

\newcommand{\conciseExpl}[1]{%
{\textit{Give Concise Explanations}}%
}

\newcommand{\singleDivExpl}[1]{%
{\textit{Explain COBOL Divisions}}%
}

\newcommand{\cobolExcitement}[1]{%
{\textit{Express Excitement For COBOL}}%
}

\newcommand{\codePurposeOverview}[1]{%
{\textit{Give Code Purpose Overview}}%
}

\newcommand{\mentionsProbVal}[1]{%
{\textit{Mention \probVal{}}}%
}

\newcommand{\cobolVarTag}[1]{%
$<$cobol\_var$>$%
}%

\newcommand{\numberTag}[1]{%
$<$number$>$%
}%

\newcommand{\cblDivTag}[1]{%
\textbf{**$<$Division\_Name$>$**:}%
}%

%% file: z-commands-imports-options/04_quotations_commands.tex
\newcommand{\personaVal}[2]{%
\colorboxBackgroundForegroundText{#1}{black}{#2}%
}%

\newcommand{\llmAdaptQuote}[6]{%
{%
    \begin{quote}%
        \small
        #1 (\personaVal{#2}{#3}): 
        ``\textit{#4}''
    \end{quote}%
}%
}%

%% file: z-commands-imports-options/05_programming_listings_commands.tex
\newcommand{\cobolCodeBlock}[3]{%
    \definecolor{codegreen}{RGB}{255 ,255 ,255}%
    \definecolor{comment}{RGB}{0, 0, 0}%
    \definecolor{string}{RGB}{215, 152, 125}%
    \definecolor{keyword}{RGB}{39, 64, 174}%
    \definecolor{codepurple}{RGB}{0, 0, 0}%
    \definecolor{backcolour}{RGB}{0, 0, 0}%
    \definecolor{varcolor}{RGB}{155, 219, 253}%
        \lstdefinestyle{CobolStyle}{%
            language = Cobol, %
            backgroundcolor=\color{white},%
            commentstyle=\color{comment},%
            keywordstyle=\color{black},%
            morekeywords = {},%
            basicstyle=\ttfamily\scriptsize,%
            showspaces=false,%
            numbers = left,
            numbersep = 4pt,
            numberstyle = \tiny \color{gray},
            breakautoindent=false,
            breakindent=0ex,
            breaklines = true,%
            showstringspaces=false,%
            showtabs=false,
            tabsize=1,%
            lineskip = -2pt
        }
        \def \colframe{lightgray}
        \def \boxrule{1pt}
    \lstset{style=CobolStyle}%
    \begin{tcolorbox}[top = 0pt, 
        bottom = 0pt, 
        left = 0pt, 
        right = 0pt, 
        colback = white, 
        sharp corners, 
        boxrule = \boxrule, 
        colframe = white,
        width = #3\linewidth%
        ]%
        \lstinputlisting[]{code_used/cobol/#2.cbl}%
        \end{tcolorbox}%
}

%% file: z-commands-imports-options/06_statistical_commands.tex
\newcommand{\ANOVA}[5]{%
(ANOVA, F(#1,#2): #3, p: $#4$, $\eta^2$: $#5$)%
}%

\newcommand{\ANOVAless}[5]{%
(ANOVA, F(#1,#2): #3, p: $<#4$, $\eta^2$: $#5$)%
}%

\newcommand{\tTest}[5]{%
(#1-sided t-test, t(#2): #3, p: $#4$, $d: #5$)%
}%